\documentclass[12pt]{article}
\usepackage{float}
\usepackage{amssymb}
\usepackage{amsthm}
\usepackage{amsmath}
\usepackage{latexsym}
\usepackage{epsfig}
\usepackage{graphicx}
\usepackage{wasysym}
\usepackage{threeparttable}
\usepackage{natbib}
\usepackage{color}
\usepackage{epstopdf}
\usepackage{caption}
\usepackage{subcaption}

\newcommand{\ind}{\stackrel{\mathrm{ind}}{\sim}}

\usepackage[breaklinks]{hyperref}

\def\boxit#1{\vbox{\hrule\hbox{\vrule\kern6pt
          \vbox{\kern6pt#1\kern6pt}\kern6pt\vrule}\hrule}}

\def\bse{\begin{eqnarray*}}
\def\ese{\end{eqnarray*}}
\def\be{\begin{eqnarray}}
\def\ee{\end{eqnarray}}
\def\bq{\begin{equation}}
\def\eq{\end{equation}}
\def\bse{\begin{eqnarray*}}
\def\ese{\end{eqnarray*}}

\usepackage{mathptmx}      
\usepackage{bm}

 {\begin{list}{}%
         {\setlength{\leftmargin}{#1}}%
         \item[]%
 }
 {\end{list}}
\usepackage{setspace}

\def\bu{\textbf{u}}
\def\bs{\textbf{s}}

\def\bmu{\bm{\mu}}

\def\bz{\textbf{Z}}

\def\bfbeta{\bm{\beta}}

\begin{document}
\thispagestyle{empty} \baselineskip=28pt

\begin{center}
{\LARGE{\bf Bayesian Spatial Change of Support for Count-Valued Survey Data}}
\end{center}

\baselineskip=12pt

\vskip 2mm
\begin{center}
Jonathan R. Bradley\footnote{(\baselineskip=10pt to whom correspondence should be addressed) Department of Statistics, University of Missouri, 146 Middlebush Hall, Columbia, MO 65211, bradleyjr@missouri.edu},
Christopher K. Wikle\footnote{\baselineskip=10pt  Department of Statistics, University of Missouri, 146 Middlebush Hall, Columbia, MO 65211-6100},
Scott H. Holan$^2$
\end{center}
%
%
%
%
\vskip 4mm

\begin{center}
\large{{\bf Abstract}}
\end{center}
We introduce Bayesian spatial change of support methodology for count-valued survey data with known survey variances. Our proposed methodology is motivated by the American Community Survey (ACS), an ongoing survey administered by the U.S. Census Bureau that provides timely information on several key demographic variables. Specifically, the ACS produces 1-year, 3-year, and 5-year ``period-estimates,'' and corresponding margins of errors, for published demographic and socio-economic variables recorded over predefined geographies within the United States. Despite the availability of these predefined geographies it is often of interest to data users to specify customized user-defined spatial supports. In particular, it is useful to estimate {demographic variables} defined on ``new'' spatial supports in ``real-time.'' This problem is known as spatial change of support (COS), which is typically performed under the assumption that the data follows a Gaussian distribution. However, count-valued survey data is naturally non-Gaussian and, hence, we consider modeling these data using a Poisson distribution. Additionally, survey-data are often accompanied by estimates of error, which we incorporate into our analysis. We interpret Poisson count-valued data in small areas as an aggregation of events from a spatial point process. This approach provides us with the flexibility necessary to allow ACS users to consider a variety of spatial supports in {``real-time.''} We demonstrate the effectiveness of our approach through a simulated example as well as through an analysis using public-use ACS data.
\baselineskip=12pt

%
%
%

\baselineskip=12pt
\par\vfill\noindent
\par\medskip\noindent
\clearpage\pagebreak\newpage \pagenumbering{arabic}
\baselineskip=24pt
\section{Introduction} Each year the U.S. Census Bureau samples approximately 3.5 million housing unit addresses through the American Community Survey (ACS). The ACS has replaced the well-known decennial Census long-form, and provides data-users with a more timely data source \citep{fey_jsm,acs}. One disadvantage of the ACS is that it provides estimates of key demographic variables based on smaller (than the decennial Census) sample sizes. Consequently, ACS estimates often have large margins of error \citep{speilman}.
 
To improve precision, the ACS produces ``period'' estimates, which are survey estimates that are aggregated over different time points to reduce the sampling variability \citep{acs}. Specifically, depending on the level of geography, the ACS produces 1-year, 3-year, and 5-year period estimates. Single year period-estimates are provided for geographic regions that consist of populations of 65,000 or larger (e.g., congressional districts and about 800 counties), 3-year period estimates are provided for geographic regions that consist of populations of 20,000 or larger (e.g., all the states, around 1,800 counties, and 900 metropolitan statistical areas), and 5-year period estimates are provided for all population sizes available (e.g., all census tracts, county subdivisions, and counties). Thus, the ACS-defined spatial support, which we call the ``source support,'' becomes finer (in general) as the period increases.

Although ACS period estimates are available on many different source supports, it would be useful to define any desired spatial support. We call a spatial support of interest, that may possibly differ from the source support, a ``target support.'' {For example, the Department of City Planning (DCP) in New York City (NYC) use ACS period estimates of demographics, and social characteristics including poverty count-data. They are interested in obtaining estimates of these variables defined on community districts (target support), but instead use aggregate census tracts (source support) \citep{dcpacs}, since ACS data are not available on NYC's community districts.}

{In general, this problem of changing the source support to the target support is known as spatial change of support (COS). Much of the current methodology is based on assuming that the underlying data is Gaussian \citep[see, e.g.,][]{GelfandZhuCarline,WileBerliner}. See \citet{Gotway} for a review. Methodologically, we} are interested in the case where the data are represented as counts, or the sum of Bernoulli events contained within a given region, which we model using the Poisson distribution. A straightforward methodology to implement in this setting is known as ``simple areal interpolation,'' where values are imputed proportionally based on the areas of each of the target's areal unit \citep{flowerden,green3,rogerskill,green2,green}. Although it is easy to implement, measures of uncertainty are not readily available and, hence, it is difficult to ascertain the performance of the estimates defined on the target support or to use the estimated quantities in an inferential setting.

To address this deficiency, \citet{mugglin98} and \citet{mugglin99} use a hierarchical Bayesian approach based on a Poisson data model. Their approach involves partitioning the source support based on the target support, and then estimating parameters defined by each set of this partition. This approach leads to extensive bookkeeping of the partitioning formed by the source support and target support, which makes implementing their Markov chain Monte Carlo (MCMC) procedure difficult. Furthermore, since the parameters of their Bayesian hierarchical model depend on the source support and choice of target support, separate MCMC procedures are required every time a new target support is proposed. Alternatively, we adopt the perspective summarized in \citet[][{Chap.} 4]{cressie-wikle-book}, where count-valued data in small areas are interpreted as an aggregation of events of a spatial point process. This approach allows us to estimate parameters at the point level and aggregate the {latent process} to any desired target support. This avoids {additional} MCMC computations if a new target support is introduced, and gives the flexibility required to analyze ACS period-estimates. For a comprehensive comparison between {spatial COS using a point process and spatial COS based on partitioning the source and target support into disjoint regions}, see \citet{gelfandCOSreview}.

Although the aforementioned point process aggregation approach is viable for count data recorded without error, there is currently no spatial COS methodology that can be applied to count-valued survey data where estimates of the variance using the survey's sampling design are considered known (e.g., see, \citet{lohr-survey} for more on sampling designs). We use this additional information to account for the variability due to survey sampling within a Bayesian hierarchical modeling (BHM) framework. This is easily done for the Gaussian spatial linear model paradigm \citep[e.g.,][]{fayherriot,aaronp_spfh} in which, the variance of an additive error term is set equal to the estimated survey variance; hence, this additive error is used to account for variability due to survey sampling. For the non-Gaussian setting this source of variability is typically ignored in the hierarchical modeling framework. Thus, an important component of the proposed BHM is that we account for the variability due to survey sampling. This general problem of spatial COS for count-valued survey data is of independent interest, and the methodology proposed in this paper can be applied to a variety of data sources in addition to ACS.

Survey sampling addresses one source of variability; however, we would also like to include the variability of the latent process and unknown parameters. To take into account the variability of the latent process and unknown parameters we use a BHM, which extends the approach of \citet{WileBerliner}. The latent process is modeled using a spatial generalized linear mixed model (GLMM) \citep[e.g., see,][]{besagYorkMollie,reich, hughes}. However, we propose an extension of the spatial random effects covariance parameter model of \citet{hughes} by incorporating {a} Givens angle prior \citep{andersonga, bergerga,kangbayes} to allow for additional complexity in the spatial covariance. {In particular, we introduce a novel Givens angle prior for areal data models.}

The remainder of this paper is organized as follows. Section 2, introduces the spatial COS methodology, which includes the proposed BHM. The data model is defined at the point level, which allows for a flexible approach to spatial COS. Additionally, this model is calibrated in a manner that incorporates known values of the survey variance. In Section 3, we demonstrate the utility of the suggested modeling approach by simulating a stratified random sample from a model that differs from the statistical model used for COS. Here, we show that one can reasonably recover the true unobserved {latent} mean-count (known from the simulation) using the methodology outlined in Section 2. {In Section 4, we then apply the COS methodology to the motivating example of ACS demographic data recorded over NYC.} Finally, concluding discussion is provided in Section 5. {For convenience of exposition, details surrounding the MCMC algorithm are left to the Appendix.}

\section{Spatial Change of Support Methodology} Consider data that are recorded at $n_{\ell}$ small areas on the $\ell$-th spatial scale $\{A_{\ell, i}: i = 1,...,n_{\ell}\}$, which creates a partitioning of the spatial domain of interest $D \subset \mathbb{R}^{d}$; that is, $D \equiv \cup_{i = 1}^{n_{\ell}}A_{\ell,i}$ and $A_{\ell,i}\cap A_{\ell,j} = \emptyset$ ($i \ne j$). Let $n \equiv \sum_{\ell}n_{\ell}$, and $\textbf{Z} \equiv (Z(A_{\ell,i}): i = 1,...,n_{\ell}, \ell = 1,...,L)^{\prime}$ be a generic $n$-dimensional vector consisting of count-valued survey data $\--$ for example, $\bz$ might consist of the number of individuals below the poverty threshold as estimated by ACS.

Notice that it is possible to observe more than one source support; that is, $\ell = 1,...,L$ where the integer $L\ge 1$. In general, the source supports $\ell = 1,...,L$ are decided \textit{a priori} by the statistical agency that releases the public-use survey data. For example, ACS provides data by census tracts and states for some variables and time periods. We order the spatial scales $\ell = 1,...,L$ from the finest resolution (e.g., $\ell = 1$ might correspond to census tracts) to the coarsest resolution (e.g., $\ell = L$ might correspond to states). Furthermore, we assume that these supports may be misaligned (or non-nested), so that $A_{m, j}$ may include all of $A_{\ell,i}$, none of $A_{\ell,i}$, or partially overlap $A_{\ell,i}$, for $i = 1,...,n_{\ell}$, $j = 1,...,n_{m}$, and $m \ne \ell$.

The goal of our analysis is to estimate the mean of $Z(\cdot)$ (denoted by $\mu(\cdot)$) evaluated at unobserved geographies $\{B_{m}:m = 1,...,M\}$ that may possibly be misaligned with the source supports; {in this context,} $\{B_{m}\}$ is called the target support. The target support is problem specific. For example, from Section 1, we see that the DCP is interested in observing ACS estimates by community districts (which we shall denote as $\{B_{m}:m = 1,...,M\}$ for this example), despite the fact that ACS does not publish these estimates.

\subsection{Spatial Change of Support} We model an element of the generic $n$-dimensional count-valued data vector $\bz$ by
	\begin{equation}\label{data:model}
	Z(A_{\ell,i})\vert \delta(\cdot), p(\cdot) \ind \mathrm{Pois}\left(\underset{A_{\ell,i}}{\int}\delta(\bu)p(\bu)d\bu\right),\hspace{5pt}i = 1,...,n_{\ell}, \ell = 1,...,L,
	\end{equation}
\noindent
where ``Pois'' stands for Poisson, $\mu(A) \equiv \underset{A}{\int}\delta(\bu)p(\bu)d\bu$, and the probability that an event is observed at point location $\bu \in D$ is $p(\bu)$. The function $\delta(\cdot)$ represents the unobserved population density in $D$, which implies that $N_{\ell,i}\equiv \underset{A_{\ell,i}}{\int}\delta(\bu)d\bu$, $i = 1,...,n_{\ell}$ and $\ell = 1,...,L$.

The key feature behind the model in (\ref{data:model}) is quantified by the term $\underset{A_{\ell,i}}{\int}\delta(\bu)p(\bu)d\bu$. Specifically, the motivating feature of (\ref{data:model}) is that count-valued data in a small area $A$ can be interpreted as an aggregation of events of an inhomogeneous spatial point process \cite[][{Chap.} 4]{cressie-wikle-book} with intensity function $\delta(\cdot)p(\cdot)$; in (\ref{data:model}), this aggregation is represented by the integral $\underset{A}{\int}\delta(\bu)p(\bu)d\bu$.

Now, it is important to note that in the applications motivating this work, survey data are not available below the first spatial resolution (geography). Hence, a necessary assumption for this setting is that for $\bs \in A_{1,i}$, $p(\bs) \equiv p_{1,i};\hspace{5pt} i = 1,...,n_{1}$. Consequently, we can write,
	\begin{equation}\label{logmean}
	Z(A_{1,i})\vert Y_{1,i} \ind \mathrm{Pois}\left(\mathrm{exp}(Y_{1,i})\right);\hspace{5pt}i = 1,...,n_{1},
	\end{equation}
\noindent
where the distribution of $\{Y_{1,i}\equiv \mathrm{log}(N_{1,i}p_{1,i})\}$, considered the ``process'' here, shall be specified in Section 2.3. 

Similar to (\ref{logmean}), one can write the distribution of $Z(A_{\ell,i})\vert \mu(A_{\ell,j})$ in terms of $\{\mu(A_{1,i})\}$ (or equivalently $\{Y_{1,i}\}$) for $j = 1,...,n_{\ell}$ and $\ell = 2,...,L$. To see this, notice that
	\begin{equation*}
	\mu(A_{\ell,j})=\underset{A_{\ell,j}}{\int}p(\bu)\delta(\bu)d\bu;\hspace{5pt}j = 1,...,n_{\ell}, \ell = 2,...,L,
	\end{equation*}
\noindent
by definition. If we partition $A_{\ell,j}$ into its potential overlap with all $A_{1,i}$ then
\begin{align}\label{partition}
\nonumber
\mu(A_{\ell,j}) &= \sum_{i = 1}^{n_{1}}\underset{A_{\ell, j} \cap A_{1,i}}{\int}p(\bu)\delta(\bu)d\bu=\sum_{i = 1}^{n_{1}}p_{1,i}\underset{A_{\ell, j} \cap A_{1,i}}{\int}\delta(\bu)d\bu\\
&=\sum_{i = 1}^{n_{1}}N_{1,i}p_{1,i}\frac{|A_{\ell, j} \cap A_{1,i}|}{|A_{1,i}|} = \sum_{i = 1}^{n_{1}}\mu(A_{\ell,i})\frac{|A_{\ell, j} \cap A_{1,i}|}{|A_{1,i}|};\hspace{5pt}j = 1,...,n_{\ell}, \hspace{5pt}\ell = 2,...,L.
\end{align}
\noindent
Let $h_{\ell}(j,i) \equiv |A_{\ell, j} \cap A_{1,i}|/|A_{1,i}|$, the $n_{1}$-dimensional vector $\textbf{h}_{\ell}(j)\equiv(h_{\ell}(j,1),...,h_{\ell}(j,n_{1}))^{\prime}$, the $n_{1}$-dimensional random vector $\bm{\mu} \equiv (\mu(A_{1,i}): i = 1,...,n_{1})^{\prime}$, and $|A|$ denote the number of elements in the set $A$. Then, from (\ref{partition}) we have
\begin{align}\label{COS}
\mu(A_{\ell,j})=  \textbf{h}_{\ell}(j)^{\prime}\bm{\mu} \hspace{2pt};\hspace{5pt}j = 1,...,n_{\ell}, \ell = 2,...,L,
\end{align}
\noindent
and
	\begin{equation}\label{logmean2}
	Z(A_{\ell,j})\vert \mu(A_{\ell,j}) \ind \mathrm{Pois}\left(\textbf{h}_{\ell}(j)^{\prime}\bm{\mu} \right);\hspace{5pt}j = 1,...,n_{\ell}, \ell = 2,...,L.
	\end{equation}
Equation (\ref{COS}) leads naturally to an approach for estimating $\{\mu(B_{m})\}$. That is, from (\ref{COS}) we have that 
\begin{align}\label{COS2}
\mu(B_{m})=  \textbf{h}_{B}(m)^{\prime}\bm{\mu} \hspace{2pt};\hspace{5pt}m = 1,...,M, \ell = 2,...,L,
\end{align}
\noindent
where $h_{B}(m,i) \equiv |B_{m} \cap A_{1,i}|/|A_{1,i}|$, the $n_{1}$-dimensional vector $\textbf{h}_{B}(m)\equiv(h_{\ell}(m,1),...,h_{\ell}(m,n_{1}))^{\prime}$, and recall $\{B_{m}\}$ is the target support. Thus, the mechanism to change the spatial support from resolutions $\ell = 1,...,L$ to the $B$-resolution is the vector $\textbf{h}_{B}(\cdot)^{\prime}$ (i.e., ``the change of support vector''), which is consistent with the traditional areal interpolation approach discussed in Section 1, but in a context that manages uncertainty. Furthermore, the product form $\textbf{h}_{B}(m)^{\prime}\bm{\mu}$ allows for straightforward posterior inference (see Section 2.6).

\subsection{Incorporating Survey Variances} Since the survey-based variance estimates of $\mathrm{var}\{Z(A_{\ell,i})\vert \mu(A_{\ell,i})\}$, denoted $\sigma_{\ell,i}^{2}$, are provided by the ACS, we would like to incorporate this information into the statistical model to improve our estimates of $\mu(A_{\ell,i})$. We take a fully Bayesian approach and assume a distribution for each $\sigma_{\ell,i}^{2}$.

{Ideally, from a ``data user" perspective the confidential microdata would be available from the statistical agency to jointly model both $\{\sigma_{\ell,i}^{2}\}$ and $\{Z(A_{\ell,i})\}$.  However, the microdata  is not made publicly available due to disclosure limitations.  Consequently, a joint model for $\{\sigma_{\ell,i}^{2}\}$ and $\{Z(A_{\ell,i})\}$ can not be specified based on the microdata. See \citet{quick_confi} and the references therein for a comprehensive discussion regarding confidentiality in the spatial setting.  } 

In general, the typical data-user has some knowledge of what to expect from the sampling distribution of $\{\sigma_{\ell,i}^{2}\}$. For example, it would be reasonable to expect: the sampling distribution of variance estimates to be unimodal and skewed right; the median of the distribution of $\sigma_{\ell,i}^{2}$ to be (hopefully) close to the true variance of $Z(A_{\ell,i})\vert \mu(A_{\ell,i})$ (i.e., $\mu(A_{\ell,i})$); and both $\sigma_{\ell,i}^{2}$ and $Z(A_{\ell,i})$ to be dependent (since they both depend on the same confidential microdata). Hence, we consider the following model for $\sigma_{\ell,i}^{2}$:
\begin{equation}\label{variances}
\mathrm{log}(\sigma_{\ell,i}^{2}) = \mathrm{log}(\mu(A_{\ell,i})) + \epsilon_{\ell,i};\hspace{5pt}i = 1,...,n_{\ell}, \ell = 1,...,L,
\end{equation}
\noindent
where $\epsilon_{\ell,i}$ are independent normal random variables with mean zero and unknown variance $\sigma_{\epsilon,\ell,i}^{2}$. Equation (\ref{variances}) implies that $\sigma_{\ell,i}^{2}$ is lognormal (and hence unimodal and skewed right) with median $\mu(A_{\ell,i})$; $i = 1,...,n_{\ell}$, and $\ell = 1,...,L$. Additionally, Equations (\ref{data:model}) and (\ref{variances}) show that both $\{Z(A_{\ell,i})\}$ and $\{\sigma_{\ell,i}^{2}\}$ are dependent on the latent process $\{\mu(A_{\ell,i})\}$. This is to be expected since $\{\sigma_{\ell,i}^{2}\}$ and $\{Z(A_{\ell,i})\}$ are dependent on the same confidential microdata, which in turn, are dependent on the latent process $\mu(A_{\ell,i})$. {Finally, (\ref{variances}) also allows our approach to be robust in cases of overdispersion (i.e., the variance is greater than the mean), through the incorporation of an extra random effect.}  Despite the simplicity of the data model (\ref{variances}), the incorporation of survey variance estimates in this way substantially improved model performance as shown in Section 3.2.

\subsection{The Process Model}Similar to \citet{WileBerliner}, we take a fully Bayesian approach to COS to account for the variability of $\{Y_{1,i}\}$. The unobserved process $\{Y_{1,i}: i = 1,...,n_{\ell}\}$ is modeled as
\begin{equation}\label{process:model:truth}
Y_{1,i}= \textbf{x}_{i}^{\prime}\bm{\beta}+\bm{\psi}_{i}^{\prime}\bm{\eta} + \xi_{i};\hspace{4pt} i = 1,...,n_{1},
\end{equation}
\noindent
where, $\textbf{x}_{i}^{\prime}\bm{\beta}$ represents ``large-scale'' variability from the $p$-dimensional known covariate vector $\textbf{x}_{\ell,i}$, and where $\bm{\beta} \in \mathbb{R}^{p}$ is the associated fixed and unknown $p$-dimensional parameter vector. The $r$-dimensional random vector $\bm{\eta}$ is assumed to be Gaussian with mean-zero and unknown precision matrix $\textbf{K}$, and $\{\xi_{i}\}$ are i.i.d. normal with mean zero and unknown variance $\sigma_{\xi}^{2}$; {note} that we reserve the word ``Gaussian'' for random vectors and the word ``normal'' for random variables.

In principle, the $r$-dimensional real vector $\bm{\psi}_{i}$ can belong to any class of spatial basis functions (e.g., EOFs, wavelets, radial basis functions, etc.); however, we shall use the Moran's I basis functions \citep[see,][]{griffith2000, griffith2002, griffith2004, griffith2007, hughes,aaronp}. The Moran's I basis functions are a class of functions used to model areal spatial processes in a reduced dimensional space (i.e., $r \ll n$). This feature allows for fast computation of the distribution of $\bm{\eta}$, which can become computationally expensive for large $r$ \citep{hughes, aaronp}. This is particularly useful for ACS data, where, for certain spatial supports (e.g., census tracts), the dataset can be quite large (on the order of thousands of areal units).

Now, $\{\bm{\psi}_{i}\}$ is defined to be the $r$ eigenvectors of the Moran's I operator \citep[see,][]{hughes} associated with the corresponding $r$ largest eigenvalues. That is, the Moran's I operator is defined as
\begin{equation*}
\textbf{M}(\textbf{A}) \equiv\left(\textbf{I}_{n_{1}} - \textbf{X}\left(\textbf{X}^{\prime}\textbf{X}\right)^{-1}\textbf{X}^{\prime}\right)\textbf{A}\left(\textbf{I}_{n_{1}} - \textbf{X}\left(\textbf{X}^{\prime}\textbf{X}\right)^{-1}\textbf{X}^{\prime}\right),
\end{equation*}
where the $n_{1}\times p$ matrix $\textbf{X} \equiv \left(\textbf{x}_{i}: i = 1,...,n_{1}\right)^{\prime}$, $\textbf{I}_{n_{1}}$ is an $n_{1}\times n_{1}$ identity matrix, and $\textbf{A}$ is the $n_{1}\times n_{1}$ adjacency matrix corresponding to the edges formed by $\{A_{1, i}: i = 1,...,n_{1}\}$. Let $\bm{\Phi}_{M}\bm{\Lambda}_{M}\bm{\Phi}_{M}^{\prime}$ be the spectral decomposition of the $n_{1}\times n_{1}$ matrix $\textbf{M}(\textbf{A})$. We follow the suggestion of \citet{hughes} and let $r$ equal (approximately) 10$\%$ of the total number of positive eigenvalues of $\textbf{M}(\textbf{A})$ (i.e., the positive diagonal elements of $\bm{\Lambda}_{M}$) and the $n\times r$ matrix $\bm{\Psi}\equiv (\bm{\psi}_{i}: i = 1,...,r)^{\prime}$ is defined to be the first $r$ columns of $\bm{\Phi}_{M}$. Specifying larger values of $r$ allows one to model higher frequencies of the hidden process (i.e., model higher resolutions of the spatial process), however, specifying a large value of $r$ also increases the time it takes to compute the distribution of $\bm{\eta}$ (e.g., see, \citet{bradley2011}, for a discussion). In practice, one needs to determine a balance between computational feasibility and model complexity.

\subsection{Parameter Models} We propose a novel extension of the spatial random effect covariance parameter model from \citet{hughes} to allow for increased flexibility. Specifically, we add structure on the prior distribution for $\textbf{K}$ (i.e., the covariance matrix of the random effect $\bm{\eta}$). \citet{hughes} set $\textbf{K} = \phi \times \bm{\Psi}^{\prime}\textbf{Q}\bm{\Psi}$, where $\phi>0$. Here, the $n_{1}\times r$ matrix $\bm{\Psi} \equiv (\bm{\psi}_{i}: i = 1,...,n_{1})^{\prime}$, the $n_{1}\times n_{1}$ matrix $\textbf{Q} \equiv \mathrm{diag}(\textbf{A}\bm{1}) - \textbf{A}$, and $\bm{1}$ is a $n_{\ell}$-dimensional vector of ones. 

Using the spectral decomposition of $\bm{\Psi}^{\prime}\textbf{Q}\bm{\Psi}$ we have $\textbf{K} = \bm{\Phi}_{Q}(\phi \times \bm{\Lambda}_{Q})\bm{\Phi}_{Q}^{\prime}$, where $\bm{\Phi}_{Q}$ is a $r\times r$ known orthogonal matrix and $\bm{\Lambda}_{Q}$ is a $r\times r$ known diagonal matrix with nonnegative entries. Hence, the model in \citet{hughes} is based on the assumption that the eigenvalues of $\textbf{K}$ are known up to a multiplicative constant and the eigenvectors of $\textbf{K}$ are known. These assumptions are rather strong since only a single parameter is used to describe the parameter space of the $r\times r$ positive-definite matrix $\textbf{K}$.

Hence, we extend this parameter model by letting $\textbf{K} = \bm{\Phi}(\phi \times \bm{\Lambda}_{Q})\bm{\Phi}^{\prime}$ and by adding a prior distribution on $\bm{\Phi}$. Specifically, we let $\bm{\Phi}$ follow a Givens Angle Prior (GAP) \citep{andersonga, bergerga,kangbayes}. That is, the $r\times r$ matrix $\bm{\Phi}$ is decomposed into a Givens rotator product,
\begin{align*}
\bm{\Phi} \equiv (\bm{O}_{1,2} \times \bm{O}_{1,3}\times\cdots \times \bm{O}_{1,r})\times (\bm{O}_{2,3} \times \cdots \times \bm{O}_{2,r})\times\cdots\times \bm{O}_{r-1,r},
\end{align*} 
\noindent
where $\bm{O}_{i,j}$ is a $r\times r$ identity matrix with the $(i,i)$-th and $(j,j)$-th element replaced by $\mathrm{cos}(\theta_{i,j})$ and the $(i,j)$-th ($(j,i)$-th) element replaced by $-\mathrm{sin}(\theta_{i,j})$ ($\mathrm{sin}(\theta_{i,j})$). Here, $\theta_{i,j} \in [-\pi/2, \pi/2]$, and we place a prior on the logit transformation of the shifted and rescaled $\theta_{i,j}$ denoted as $\zeta_{i,j} \equiv 1/2 + \theta_{i,j}/\pi$. That is,
\begin{equation}\label{logit}
\mathrm{logit}(\zeta_{i,j}) = a + b \times g_{i,j}(\bm{\Phi}_{Q});\hspace{4pt}i<j = 1,...,r,
\end{equation}
\noindent
where $\mathrm{logit}(\zeta_{i,j}) \equiv \mathrm{log}\{\zeta_{i,j}/(1-\zeta_{i,j})\}$, $a,b \in \mathbb{R}$, and $g_{i,j}$ represents the $(i,j)$-th Givens angle of $\bm{\Phi}_{Q}$ (e.g., see \citet{bergerga} for more details on the Givens angle). We then place a vague Gaussian prior on $(a,b)^{\prime}$. For the remaining parameters we also assume vague prior distributions. Specifically, we assume $\bm{\beta}$ has a vague Gaussian prior, and $\phi$, $\sigma_{\gamma}^{2}$, and $\sigma_{\epsilon,\ell,i}^{2}$ ($\ell = 1,...,L$ and $i = 1,...,n_{\ell}$) are each assumed to have vague inverse gamma (IG) priors.

Typically, one places independent Gaussian priors on each $\mathrm{logit}({\zeta_{i,j}})$ \citep{andersonga, bergerga,kangbayes}. However, this would lead to a total of $r \times (r-1)/2$ parameters; hence, for moderately large values of $r$ this would lead to estimating many parameters. For example, if $r=85$ then there are a total of $85\times 42=3,570$ parameters that define the eigenvectors of the covariance matrix $\textbf{K}$. For the parameter model (\ref{logit}), there are only 2 parameters.

\subsection{A Summary of the Bayesian Hierarchical Model} The joint distribution of the data, processes, and parameters is written as the product of the following conditional distributions:
\begin{subequations}
\begin{align}
\nonumber
&\mathrm{Data\hspace{5pt}Model\hspace{5pt}1:}\hspace{5pt} Z(A_{\ell,i})\vert \bm{\eta}, \bfbeta, \xi_{i}\ind \mathrm{Pois}\{\mu(A_{\ell,i})\};\\
\label{summarya}
&\mathrm{Data\hspace{5pt}Model\hspace{5pt}2:}\hspace{5pt} \mathrm{log}(\sigma_{\ell,i}^{2})\vert \bm{\eta}, \bfbeta, \xi_{i}\ind \mathrm{Normal}\{\mathrm{log}(\mu(A_{\ell,i})), \sigma_{\epsilon,\ell,i}^{2}\};\\
\nonumber
&\mathrm{Process\hspace{5pt}Model\hspace{5pt}1:}\hspace{5pt} \bm{\eta}\vert \bfbeta, (a, b)^{\prime}, \phi \sim \mathrm{Gaussian}\left\lbrace\bm{0}, \textbf{K}(a,b,\phi)\right\rbrace;\\
\nonumber
&\mathrm{Process\hspace{5pt}Model\hspace{5pt}2:}\hspace{5pt} \xi_{j}\vert \sigma_{\gamma}^{2} \sim \mathrm{Normal}\left(0, \sigma_{\gamma}^{2}\right);\\
\nonumber
&\mathrm{Parameter\hspace{5pt}Model\hspace{5pt}1:}\hspace{5pt} \bfbeta \sim \mathrm{Gaussian}\left(\bm{\mu}_{\beta}, \sigma_{\beta}^{2}\textbf{I}_{p}\right);\\
\label{summaryb}
&\mathrm{Parameter\hspace{5pt}Model\hspace{5pt}2:}\hspace{5pt}  \phi \sim \mathrm{IG}(\alpha_{\phi}, \omega_{\phi});\\
\label{summaryc}
&\mathrm{Parameter\hspace{5pt}Model\hspace{5pt}3:}\hspace{5pt} (a,b)^{\prime} \sim \mathrm{Gaussian}\{\bm{\mu}_{\Phi}, \sigma_{\Phi}^{2}\textbf{I}_{2}\};\\
\label{summaryd}
&\mathrm{Parameter\hspace{5pt}Model\hspace{5pt}4:}\hspace{5pt}  \sigma_{\epsilon,\ell,i}^{2} \sim \mathrm{IG}(\alpha_{\epsilon}, \omega_{\epsilon})\\
\label{summarye}
&\mathrm{Parameter\hspace{5pt}Model\hspace{5pt}5:}\hspace{5pt}  \sigma_{\gamma}^{2} \sim \mathrm{IG}(\alpha_{\gamma}, \omega_{\gamma}),
\end{align}
\end{subequations}
\noindent
where $\ell = 1,...,L, i = 1,...,n_{\ell}$, and $j = 1,...,n_{1}$. In practice, to use this algorithm we have to specify values for the hyperparameters $\bm{\mu}_{\beta}$, $\sigma_{\beta}^{2}$, $\alpha_{\phi}$, $\omega_{\phi}$, $\mu_{\Phi}$, $\sigma_{\Phi}^{2}$, $\alpha_{\epsilon}$, $\omega_{\epsilon}$, $\alpha_{\gamma}$, and $\omega_{\gamma}$. We choose values for these hyperparameters so that the corresponding prior distribution is flat and, hence, relatively noninformative. Specifically, we let $\bm{\mu}_{\beta} = \bm{0}$, $\sigma_{\beta}^{2} = 10^{15}$, $\alpha_{\phi} = \omega_{\phi}= 1$, $\bmu_{\Phi} = (0,1)^{\prime}$, $\sigma_{\Phi}^{2} = 10^{15}$, $\alpha_{\epsilon}=\alpha_{\gamma} = 1$, and $\omega_{\epsilon} = \omega_{\gamma} = 1$. From our experience, this procedure is rather robust to changes in the prior variance values (i.e., changing $10^{15}$ to $1000$); however, performance tends to be sensitive to choices of $\bmu_{\Phi}$. In particular, the procedure performs well when $\bmu_{\Phi}$ is specified close to $(0,1)^{\prime}$, and appears to do worse as $\bmu_{\Phi}$ is set to values much different than $(0,1)^{\prime}$. {The value $\bmu_{\Phi} = (0,1)^{\prime}$ corresponds to a standardized CAR model.}
  
  Now, using the Metropolis-Hastings algorithm given in the Appendix, we obtain samples from the posterior distributions of $\bm{\beta}$, $\bm{\eta}$, and $\{\xi_{i}\}$, which we denote with $\bm{\beta}_{k}^{*}\equiv (\beta_{k,1}^{*},...,\beta_{k,p}^{*})^{\prime}$, $\bm{\eta}_{k}^{*}\equiv (\eta_{k,1}^{*},...,\eta_{k,r}^{*})^{\prime}$, and $\{\xi_{i,k}: i = 1,...,n_{1}, k = 1,....,K\}$, respectively; $k = 1,...,K$. In general, we let the superscript ``*'' denote a sample from the posterior distribution. Posterior samples of $\mu(A_{1,i})$ and $\bm{\mu}$ are given by $\mu_{k}^{*}(A_{1,i}) \equiv \mathrm{exp}\left(\textbf{x}_{i}^{\prime}\bm{\beta}_{k}^{*}+\bm{\psi}_{i}^{\prime}\bm{\eta}_{k}^{*}+ \xi_{i,k}\right)$ and $\bm{\mu}_{k}^{*} \equiv (\mu_{k}^{*}(A_{1,i}): i = 1,...,n_{1})^{\prime}$, respectively; $i = 1,...,n_{1}$ and $k = 1,...,K$. The result in (\ref{COS2}) makes it straightforward to produce estimates of $\mu(\cdot)$ defined on the target support $\{B_{m}\}$; that is, from (\ref{COS2}), $\mu_{k}^{*}(B_{m}) = \textbf{h}_{B}(m)^{\prime}\bm{\mu}_{k}^{*}$, where $m = 1,...,M$ and $k = 1,...,K$. Once one obtains the posterior samples $\{\mu_{k}^{*}(B_{m})\}$, any statistic can be computed from these samples to make inference on $\mu(B_{m})$ for $m = 1,...,M$.
  
  There are other types of statistical inference that are of interest besides estimates of $\mu(\cdot)$. For example, in Section 5, we use the posterior predictive $p$-value \citep{rubinpostpval,mengpostpval,gelmanpostpval} to assess the {goodness-of-fit} of the model. To do this, replicates from the posterior predictive distribution are found by generating $K$ replicates from the distribution of $\bz, \{\sigma_{\ell,i}^{2}\}\vert \bm{\eta}_{k}^{*},\bm{\beta}_{k}^{*},\bm{\epsilon}_{k}^{*}$, where $k = 1,...,K$. 
  
  Using the $K$ replicates from $\bz, \{\sigma_{\ell,i}^{2}\}\vert \bm{\eta}_{k}^{*},\bm{\beta}_{k}^{*},\bm{\epsilon}_{k}^{*}$, denoted $\{\bz_{k} ^{*}, \{\sigma_{\ell,i}^{2}\}_{k}^{*}: k = 1,...,K\}$, we compute the likelihood ratio to evaluate if the observed dataset {represents} a typical value from the posterior predictive distribution. Specifically, we compute
  \begin{equation}\label{post:pval}
  \frac{[\bz_{k}^{*}, \{\sigma_{\ell,i}^{2}\}_{k}^{*}\vert \bm{\eta}_{k}^{*},\bm{\beta}_{k}^{*},\bm{\epsilon}_{k}^{*},\phi_{k}^{*},\bm{\Phi}_{k}^{*}]}{[\bz, \{\sigma_{\ell,i}^{2}\}\vert \bm{\eta}_{k}^{*},\bm{\beta}_{k}^{*},\bm{\epsilon}_{k}^{*},\phi_{k}^{*},\bm{\Phi}_{k}^{*}]};\hspace{5pt}k = 1,...,K,
  \end{equation}
  \noindent
  where, in general, $[X\vert Y]$ denotes the p.d.f. of the conditional distribution of $X$ given $Y$. Then, the posterior predictive $p$-value is computed as the proportion of $K$ likelihood ratios in (\ref{post:pval}) that are greater than one. Posterior predictive $p$-values very close to zero (one) indicate that the model is producing estimates that are not (too) ``close'' to the data. Hence, posterior predictive $p$-values near 0.5 are preferable.

\section{Simulations} In this section, a simulated example is used to demonstrate that one can accurately estimate the true values of $\{\mu_{\ell,i}\}$ and $\{\mu(A_{\ell+1,j})\}$ using the proposed methodology. Additionally, a simulation study is used to determine: whether it is useful to incorporate estimates of survey variances using (\ref{variances}); whether the Givens angle prior leads to improvements in prediction; and to assess the computational feasibility of the proposed method for spatial COS. To do this, we generate data from a pre-specified survey design that differs from the proposed model. This approach results in simulated data that more closely resemble a survey that one might observe in practice. 

We begin by generating a collection of ``pseudo-households'' denoted by the set $\{\bu_{j}: j = 1,...,N\}$ - by ``pseudo-household,'' we mean that in a specific context $\{\bu_{j}: j = 1,...,N\}$ might represent all the households in some specified geographical region. We generate the set $\{\bu_{j}: j = 1,...,N\}$ by performing the following steps.
\begin{enumerate}
\item Divide the square $D\equiv \{\bu: \bu = (u_{1},u_{2})^{\prime}, u_{1}\in [1,12],u_{2}\in [1,12]\} \subset \mathbb{R}^{2}$ into a $6\times 6$ grid (Figure 1). Then randomly select three grid-cells (using a discrete uniform distribution) to represent ``hot spots,'' which are colored black.
\item For each grid-cell $j$ that is a hot spot (i.e., colored black) sample $5000$ points uniformly within a circle with radius two and center equal to the center of the $j$-th grid-cell. (Note, there are three such grid cells in Figure 1.)
\item For each grid-cell $j$ that is not a hot spot (i.e., colored white) sample $1000$ points uniformly within the $j$-th grid-cell.
\end{enumerate}
\noindent
Notice that the set $\{\bu_{j}: j = 1,...,N\}$ is generated from a survey-design that differs from the model that we use to fit the data to better match the type of survey data encountered by a typical user that does not have access to microdata. 

Each ``pseudo-household'' in the set $\{\bu_{j}: j = 1,...,N\}$ references a dichotomous variable denoted $\{w(\bu_{j}): j = 1,...,N\}$. For example, if $\{\bu_{j}: j = 1,...,N\}$ represents each household in the geographic region of interest, then $\{w(\bu_{j}): j = 1,...,N\}$ consists of zeros and ones, where $w(\bu_{j})=1$ ($w(\bu_{j})=0$) could indicate that the family that resides in household $j$ falls below (above) the poverty line. To generate the values for $w(\cdot)$, we let the outcome $w(\bu)=1$ have probability 0.5 for each $\bu \in \{w(\bu_{j}): j = 1,...,N\}$.

Suppose a statistical agency would like to collect a stratified random sample of $\{w(\bu_{j}): j = 1,...,N\}$. In practice, a statistical agency would likely use a more complex survey-design, but a stratified random sample is enough for the purposes of this simulation study. Divide the region $D$ into the collection of areal units $\{A_{1,i}: i = 1,...,90\}$ displayed in Figure 2(a). Here, the areal units $\{A_{1,i}: i = 1,...,90\}$ represent ``strata.'' In practice, it is not always possible to observe $w(\cdot)$ at all $N$ households. Thus, a sample of size 50 is taken from each areal unit in the set $\{A_{1,i}\}$, which we denote as $\{\bs_{i,j}: i = 1,...,90, i = 1,...,50\} \subset \{\bu_{j}:j = 1,...,N\}$. In Figure 2(b), we display $\{\bs_{i,j}\}$ and indicate where $w(\bs) = 1$ (or $= 0$) for each $\bs \in \{\bs_{i,j}\}$. 

Based on the strata in Figure~2(a), the statistical agency then produces estimates of $\mu(A_{\ell,i}) = \underset{\bu\in A_{\ell,i}}{\sum}w(\bu)$; notice that from the simulation we know the values of $\mu(\cdot)$. For this simulated example, the statistics used for stratified random samples \citep[][pg. 99$\--$102]{lohr-survey} are computed as:
\begin{equation}\label{survey:weights}
Z(A_{1,i}) = \frac{N_{1,i}}{50} \underset{\bs \in A_{1,i}}{\sum} w(\bs);\hspace{5pt} i = 1,...,90,
\end{equation}
\begin{equation}\label{variance:survey:weights}
\sigma_{1,i}^{2} = \left(1 - \frac{50}{N_{1,i}}\right)N_{1,i}^{2} \frac{\widehat{p}_{1,i}(1-\widehat{p}_{1,i})}{49};\hspace{5pt} i = 1,...,90.
\end{equation}
\noindent
Here, $\widehat{p}_{1,i} \equiv \underset{\bs \in A_{1,i}}{\sum} w(\bs) / 50$ is the sample proportion, and $N_{1,i}$ represents the number of elements in the set $\{\bu_{j}: j = 1,...,N\}\cap A_{1,i}$; $i = 1,...,n$. In Figures~2(c) and 2(d), we display $\{Z(A_{\ell,i})\}$ and $\{\sigma_{\ell,i}^{2}\}$ computed using the realized values of $\{w(\bs_{i,j}): i = 1,...,90, j = 1,...,50\}$ in Figure 2(b) and the strata in Figure~2(a).

\subsection{A Simulated Example}

Consider a data-user who is interested in obtaining estimates and measures of uncertainty defined on a different support than what is provided by the statistical agency in Figures~2(c)-2(d). Specifically, consider the target support presented in Figure~2(e) which, as seen in Figure~2(f), is misaligned with the source support $\{B_{m}: m = 1,...,36\}$. To use the proposed model to solve this problem we need to define $\{\textbf{x}_{i}\}$, $r$, and prior distribution hyperparameters. Here, we take $\textbf{x}_{i}\equiv 1$, we let $r$ equal roughly $10\%$ of the number of positive eigenvalues of the Moran's I operator, and set the prior parameters equal to the choices given in Section~2.6. The MCMC algorithm in the Appendix was run for 15,000 iterations with a burn-in of 5,000 iterations. Convergence of the MCMC algorithm was assessed visually using trace plots {of the sample chains, with no lack of convergence detected.}

Figure~3(a,c,d,f) displays the posterior mean and posterior variance for $\{\mu(A_{1,i})\}$ and $\{\mu(B_{m})\}$, respectively. When comparing the estimates (Figure~3(a, d)) to the truth (Figure~3(b, e)), we see that the posterior mean reproduces the general trend of the true values of $\{\mu(A_{1,i})\}$ and $\{\mu(B_{m})\}$. That is, larger values of the posterior mean and larger values of the truth are located near the black grid-cells displayed in Figure 1. This is also seen in Figure~3(g,h), where we plot $\{\mu(A_{\ell,i})\}$ and $\{\mu(B_{m})\}$ and the corresponding posterior means. Hence, it appears that the estimates of $\{\mu(A_{\ell,i})\}$ and $\{\mu(B_{m})\}$ are close to their respective true values.

\subsection{Multiple Simulated Replicates} The BHM in (11) is currently the only stochastic modeling approach for spatial change of support of count-valued survey data (henceforth referred to as ``CS'' for count-valued survey-data). In contrast, the primary (non-stochastic) modeling approach that one can directly apply to survey data is simple areal interpolation (SI). Thus, we shall compare the CS model to SI. 

We also consider comparing to straightforward adjustments to the BHM in (11) to provide motivation for the data model in (\ref{variances}) and the parameter models in (\ref{summaryb}) $\--$ (\ref{summarye}). In particular, we assert that using the data model in (\ref{variances}) leads to more precise estimates than simply ignoring the survey variances $\{\sigma_{\ell,i}^{2}\}$. Hence, we consider an alternative model that removes (\ref{summarya}) from the BHM in (11). We shall refer to this BHM as the variance removed (VR) model. Additionally, a Givens angle prior is proposed to provide greater flexibility for modeling spatial covariances. To determine whether this added model complexity is beneficial we compare to a BHM that uses the Moran's I prior instead. Specifically, we replace the parameter models in (\ref{summaryb}) $\--$ (\ref{summarye}) with
\begin{equation*}
\phi \sim \mathrm{IG}(\alpha_{\phi}, \gamma_{\phi}),
\end{equation*}
\noindent
where $\textbf{K} = \phi\times \bm{\Psi}^{\prime}\textbf{Q}\bm{\Psi}$. We shall refer to this BHM as the MI model. In what follows, we compare the CS, VR, MI, and SI over 50 replicates of the simulated field $\{Z(A_{\ell,i})\}$.

To evaluate these models over 50 replicates of $\bz$ we define the \textit{paired absolute deviation} (PAD),
\begin{equation}\label{cht4.response}
\mathrm{PAD}(\mathrm{MD},\bz) \equiv \frac{1}{M}\sum_{m=1}^{M}\mathrm{abs}\left\lbrace \mu(B_{m}) - E_{\mathrm{MD}}(\mu(B_{m})\vert \bz)\right\rbrace - \frac{1}{M}\sum_{m=1}^{M}\mathrm{abs}\left\lbrace \mu(B_{m}) - E_{\mathrm{CS}}(\mu(B_{m})\vert \bz)\right\rbrace,
\end{equation}
\noindent
where the function ``abs'' is the absolute value function, $E_{\mathrm{MD}}(\mu(B_{m})\vert \bz)$ is the posterior expectation of $\mu(B_{m})$ with respect to model MD, and MD = VR, MI, and SI. Recall that $\{\mu(B_{m}): m = 1,...,36\}$ in (\ref{cht4.response}) is known from the simulation. If PAD in (\ref{cht4.response}) is positive for a given replicate of the field $\{Z(A_{1,i}): i = 1,...,80\}$, then the CS based estimate of  $\{\mu(B_{m}): m = 1,...,36\}$ is considered ``better than'' the MD based estimate.

Thus, PAD in (\ref{cht4.response}), can be considered the response in a paired experiment between MD (= VR, MI, and SI) and the CS based estimate of $\{\mu(B_{m}): m = 1,...,36\}$. Boxplots of PAD in (\ref{cht4.response}) over 50 replicates of $\{Z(A_{1,i}): i = 1,...,80\}$ are given in Figure 4(a). Notice that the boxplots corresponding to MD = VR and MI are almost completely greater than zero indicating that our two modeling decisions (i.e., incorporating survey variances and extending the MI prior) leads to improvements in spatial change of support. Also, the boxplot corresponding to SI is completely above zero, which indicates that we obtain a marked improvement by using CS over the benchmark SI. These results are further corroborated with $p$-values from sign-tests, which are 4.53$\times 10^{-14}$, 4.53$\times 10^{-14}$, and 8.88$\times 10^{-16}$ for MD = VR, MI, and SI, respectively.

When comparing the three boxplots in Figure~4(a) we also see that MD = VR and MI do not lead to substantial differences between the PAD of SI; specifically, the medians across the different values for MD are similar, and the IQRs are relatively large. Hence, not only do we see that CS leads to improvements over SI, but if we do not incorporate the survey variances and use the extended version of the MI prior, then the estimates of $\{\mu(B_{m}): m = 1,...,36\}$ do not appear to be substantially different from those generated from SI.

Recall that we use the change of support methodology given in Section 2.1 because it allows for efficient spatial change of support, and it avoids computationally expensive MCMC calculations each time a new target support is introduced. Here, we provide CPU times for the MCMC algorithm and for spatial COS using Equation (\ref{COS2}) to support this claim. {All computations were carried out in Matlab using a Dell Optiplex 7010 Desktop Computer with a quad-Core 3.40 GHz processor and 8 Gbytes of memory.} The CPU time involved with MCMC calculations and spatial COS, for each of the 50 replicates of $\bz$, is summarized in the histograms given in Figure 4~(b,c). These plots illustrate that the MCMC algorithm takes a moderate amount of time (the median is approximately 93.03 seconds), and spatial change of support can be computed in real time; that is, the CPU times are consistently close to the median CPU time given by approximately 0.03 seconds.

\section{Application: Poverty in New York City} The Department of City Planning (DCP) in New York City (NYC) is involved in promoting growth in the communities of NYC. They achieve this, in part, by establishing policies and zoning changes for the entire city and for individual neighborhoods. The DCP also provides technical support related to housing, public transit, public space, and community facilities.

The period estimates from ACS are useful for DCP's mission \citep{dcpacs}. In particular, they use ACS period estimates of demographics, and social characteristics. In this example, we consider the 2012 5-year period estimates of the population below the poverty threshold in NYC. These estimates and their corresponding estimated survey variances are displayed in Figure~5. The scatterplot of log$(\sigma_{\ell,i}^{2})$ versus log$(Z(A_{\ell,i}))$ indicates overdispersion. Hence, a Poisson model with random effects appears reasonable for this dataset (e.g., see \citealp[][pg. 198-199]{glm-nelder}, on using random effects to allow for overdispersion in Poisson models). The source support for this example is given by census tracts (Figure~6), and is denoted by $\{A_{1,1},...,A_{1,2166}\}$.

It is of interest to observe the poverty estimates referenced by community districts, which we denote with $\{B_{1},...,B_{71}\}$; see Figure~6 for the boundaries of the community districts in NYC. However, ACS period estimates are not available by community districts. In Figure~7, we see that public use microdata areas (PUMA) (aggregate census tracts) are not coterminous with community districts. Despite this misalignment, the DCP treats PUMAs as an approximation to community districts and suggests using PUMAs in place of community districts \citep{dcpacs}. This simple approximation can be done using SI, which is presented in Figure~7.

Instead of compromising by using PUMAs to deal with the misaligned spatial supports, we use the proposed spatial COS methodology to change the spatial support of the 2012 5-year period estimates of poverty from census tracts (source support) to community districts (target support). In Figure~8, we provide the posterior means and variances by census tracts and community districts. These predictions are based on the methodology proposed in Section~2 with $\textbf{x}_{\ell,i}\equiv 1$, $r=85$, and we set the prior parameters equal to the choices given in Section~2.6. The maximum value of $r$ is 844 and, following the advice of \citet{hughes}, we let $r$ equal roughly 10$\%$ of the available Moran's I basis functions {yielding $r = 85$}. The MCMC algorithm in the Appendix was run for 20,000 iterations with a burn-in of 10,000 iterations. Convergence of the MCMC algorithm was assessed visually using trace plots {of the sample chains, with no lack of convergence detected}.

Comparing the model based predictions in Figure~8 to the approximation suggested by DCP presented in Figure~7, we see that the model based estimates of $\{\mu(B_{m})\}$ share the same pattern as the DCP approximations. Hence, we appear to be obtaining reasonable estimates of $\{\mu(B_{m})\}$ using the model in Section~2. Additionally, notice in Figure 8 that the model based predictions defined on the census tracts ``smooth'' the observed dataset. That is, the larger values of our predictions correspond to the larger values within the dataset, but not quite as large. We would like to determine if we have oversmoothed (or undersmoothed) the data. To do this, we compute the posterior predictive $p$-value (see Section~2.6). Here, the posterior predictive $p$-value is 0.60, which indicates {no lack of fit; that is,} we are obtaining a reasonable fit to the data \citep{rubinpostpval,mengpostpval,gelmanpostpval} on the source support (i.e., census tracts).

\section{Discussion} We have developed a spatial change of support (COS) methodology to analyze count-valued {survey} data with known survey variances. The COS problem is typically solved under the assumption that the data follows a Gaussian distribution \citep[e.g.,][]{GelfandZhuCarline,Gotway,WileBerliner}. However, count-valued data is naturally non-Gaussian and, hence, we extend the Gaussian spatial COS methodology to Poisson spatial COS methodology. 

There are two key motivating features behind the proposed statistical model. The first motivating feature, is that count-valued data in a small area $A$ can be interpreted as an aggregation of events of an inhomogeneous spatial point process \cite[][{Chap.} 4]{cressie-wikle-book}. Hence, parameters are estimated at the point level, which allows one to aggregate the latent process to any desired target support. This avoids implementation of computationally expensive MCMC algorithms every time a target support is introduced, which are required by other spatial COS methodologies that use the Poisson distribution \citep{mugglin98,mugglin99}. The second motivating feature of the proposed approach is that we include a random component to account for the variability due to survey sampling. This allows us to account for much of the variability due to survey sampling, which for the ACS data, can be quite large \citep{speilman}.

In addition to incorporating survey sampling error, we also introduce a BHM that accounts for the variability of the latent process and the parameters. The latent process is modeled using a spatial generalized linear mixed model (GLMM) framework that includes a spatial random effect \citep[e.g., see,][]{besagYorkMollie,reich, hughes}. We also propose an extension of the spatial random effects covariance parameter model from \citet{hughes}, which allows for additional model complexity in the spatial covariance.

A simulated example is presented in which we simulate data at the point level and use a stratified random sample to produce survey-based estimates. Although the spatial COS methodology differed from the simulation model, it is apparent from the results of the simulation that one can accurately estimate the unobserved (``true'') mean using the proposed methodology.

A real data analysis based on demographic data obtained from the American Community Survey (ACS) is used to illustrate the COS methodology. In this example, we observe the estimated population below the poverty level reported at New York City's census tracts. We use the proposed spatial COS methodology to change the spatial support from census tracts to community districts. Then, the posterior predictive $p$-value is used to evaluate the performance of the estimated mean population below the poverty level. The results of the {application} suggest that we appear to be obtaining reasonable estimates using the proposed methodology.

The BHM defined in Section~2 uses a reduced rank modeling approach by extending the parameter model from \citet{hughes}. We have chosen to take the suggestion of \citet{hughes} and use columns that correspond to 10$\%$ of the total number of positive eigenvalues. One could make the model more complex and use a stochastic search variable selection (SSVS) algorithm (e.g., see \citealp{wikle-holan}) to choose spatial basis functions. However, the results of our simulation and application showed that this level of model complexity was not necessary.

There are natural avenues for future research building on the proposed methodology. For example, it would be straightforward to include exogenous covariate information within the large-scale variability term of the model. This would allow one to use this BHM for small area estimation problems. Also, there is a natural extension of the spatial random effect $\bm{\eta}$ to a spatio-temporal random effect $\bm{\eta}_{t}$, where $t$ indexes each year within a given time-period. This generalization would involve both a spatial and a temporal change of support {and is a subject of future research.}

Another possible extension of the application of our BHM would be downscaling; notice there are no restrictions on $\{B_{m}\}$ to be coarser than $\{A_{1,i}\}$. However, we note that adjustments to our methodology may be necessary, since there is no information available below $\{A_{1,i}\}$. One interesting route would be to incorporate fine-scale information about the process by replacing $\textbf{Q}$ (in Section 3) with a matrix built from a deterministic model (if available).

\section*{Acknowledgments} We would like to thank two anonymous referees, the associate editor, and the editor for their helpful comments. This research was partially supported by the U.S. National Science Foundation (NSF) and the U.S. Census Bureau under NSF grant SES-1132031, funded through the NSF-Census Research Network (NCRN) program.

\bibliographystyle{jasa}  
\bibliography{myref}

\section*{Appendix: Bayesian inference using Markov chain Monte Carlo}
\renewcommand{\theequation}{A.\arabic{equation}}
\setcounter{equation}{0}
 The MCMC procedure can be implemented using a Metropolis-Hastings within Gibbs sampler with updates as follows. The model in (11) yields,
 \begin{align}\label{joint}
 \nonumber
 &[\bm{\beta},\bm{\eta},\{\xi_{i}\}, \phi, (a,b)^{\prime}, \{\sigma_{\epsilon,\ell,i}^{2}\}, \sigma_{\gamma}^{2}\vert \bz, \{\sigma_{\ell,i}^{2}\}] \\
              \nonumber
 & \propto \hspace{4pt}\mathrm{exp}\left(\sum_{\ell = 1}^{L}\sum_{i = 1}^{n_{\ell}}Z(A_{\ell,i})Y_{\ell,i} - \mathrm{exp}(Y_{\ell,i})\right)\times \mathrm{exp}\left(-\sum_{\ell = 1}^{L}\sum_{i=1}^{n_{\ell}}\frac{(\mathrm{log}(\sigma_{\ell,i}^2) - Y_{\ell,i})^{2}}{2\sigma_{\epsilon,\ell,i}^{2}}\right)\\
 \nonumber
  & \times \mathrm{exp}\left(-\frac{\bm{\eta}^{\prime}\textbf{K}^{-1}\bm{\eta}}{2}\right) \times \mathrm{exp}\left(-\sum_{i=1}^{n_{1}}\frac{\xi_{i}^{2}}{2\sigma_{\gamma}^{2}}\right)\\
  \nonumber
 &\times \mathrm{exp}\left(-\frac{(\bm{\beta} - \bm{\mu}_{\beta})^{\prime}(\bm{\beta} - \bm{\mu}_{\beta})}{2\sigma_{\beta}^{2}}\right) \times  \phi^{-\alpha_{\phi}-1} \times \mathrm{exp}\left(-\frac{\omega_{\phi}}{\phi}\right) \\
  \nonumber
 &\times \mathrm{exp}\left(-\frac{((a,b)^{\prime} - \bm{\mu}_{\Phi})^{\prime}((a,b)^{\prime} - \bm{\mu}_{\Phi})}{2\sigma_{\Phi}^{2}}\right)\\
 \nonumber
 &\times \left\lbrace \prod_{\ell = 1}^{N}\prod_{i = 1}^{n_{\ell}}(\sigma_{\epsilon,\ell,i}^{2})^{-\alpha_{\epsilon}-1}\right\rbrace \times \mathrm{exp}\left(-\omega_{\epsilon}\sum_{\ell = 1}^{L}\sum_{i=1}^{n_{\ell}}\frac{1}{\sigma_{\epsilon,\ell,i}^{2}}\right)\\
  &\times (\sigma_{\gamma}^{2})^{-n_{1}\alpha_{\epsilon}-n_{1}} \times \mathrm{exp}\left(-\frac{n_{1}\omega_{\gamma}}{\sigma_{\gamma}^{2}}\right)
 \end{align}
 \noindent
{Now, organize the processes and parameters into a set $\bm{\theta} \equiv \{\bm{\beta},\bm{\eta},\{\xi_{i}\}, \phi, (a,b)^{\prime}, \{\sigma_{\epsilon,\ell,i}^{2}\}, \sigma_{\gamma}^{2}\}$. Let $\bm{\theta}_{-\eta} \equiv \bm{\theta}\cap\{\bm{\eta}\}^{c}$, where ``$c$'' is the set complement; similarly, one can define $\bm{\theta}_{-\beta}$, $\bm{\theta}_{-\xi}$, $\bm{\theta}_{-\epsilon,i}$, $\bm{\theta}_{-(a,b)}$, $\bm{\theta}_{-\phi}$, and $\bm{\theta}_{-\gamma}$. Some of the corresponding full-conditionals can not be sampled directly. Hence, from (\ref{joint}) we have
  \begin{align}
  \nonumber
 [\bm{\eta}\vert \bm{\theta}_{-\eta},\bz, \{\sigma_{\ell,i}^{2}\}] \propto & \hspace{4pt}\mathrm{exp}\left(\sum_{\ell = 1}^{L}\sum_{i = 1}^{n_{\ell}}Z(A_{\ell,i})Y_{\ell,i} - \mathrm{exp}(Y_{\ell,i})\right)\times \mathrm{exp}\left(-\sum_{\ell = 1}^{L}\sum_{i=1}^{n_{\ell}}\frac{(\mathrm{log}(\sigma_{\ell,i}^2) - Y_{\ell,i})^{2}}{2\sigma_{\epsilon,\ell,i}^{2}}\right)\\
   \nonumber
    &\times \mathrm{exp}\left(-\frac{\bm{\eta}^{\prime}\textbf{K}^{-1}\bm{\eta}}{2}\right),\\
   \nonumber
 [\bm{\beta}\vert \bm{\theta}_{-\beta},\bz, \{\sigma_{\ell,i}^{2}\}] \propto  & \hspace{4pt}\mathrm{exp}\left(\sum_{\ell = 1}^{L}\sum_{i = 1}^{n_{\ell}}Z(A_{\ell,i})Y_{\ell,i} - \mathrm{exp}(Y_{\ell,i})\right)\times \mathrm{exp}\left(-\sum_{\ell = 1}^{L}\sum_{i=1}^{n_{\ell}}\frac{(\mathrm{log}(\sigma_{\ell,i}^2) - Y_{\ell,i})^{2}}{2\sigma_{\epsilon,\ell,i}^{2}}\right)\\
              \nonumber
      &\times \mathrm{exp}\left(-\frac{(\bm{\beta} - \bm{\mu}_{\beta})^{\prime}(\bm{\beta} - \bm{\mu}_{\beta})}{2\sigma_{\beta}^{2}}\right)\\
         \nonumber
 [\{\xi_{i}\}\vert \bm{\theta}_{-\xi},\bz, \{\sigma_{\ell,i}^{2}\}] \propto & \hspace{4pt}\mathrm{exp}\left(\sum_{\ell = 1}^{L}\sum_{i = 1}^{n_{\ell}}Z(A_{\ell,i})Y_{\ell,i} - \mathrm{exp}(Y_{\ell,i})\right)\times \mathrm{exp}\left(-\sum_{\ell = 1}^{L}\sum_{i=1}^{n_{\ell}}\frac{(\mathrm{log}(\sigma_{\ell,i}^2) - Y_{\ell,i})^{2}}{2\sigma_{\epsilon,\ell,i}^{2}}\right)\\
              \nonumber
            &\times \mathrm{exp}\left(-\sum_{i=1}^{n_{1}}\frac{\xi_{i}^{2}}{2\sigma_{\gamma}^{2}}\right)\\
          \nonumber
 [(a,b)^{\prime}\vert \bm{\theta}_{-(a,b)},\bz, \{\sigma_{\ell,i}^{2}\}]  \propto & \hspace{4pt} \mathrm{exp}\left(-\frac{((a,b)^{\prime} - \bm{\mu}_{\Phi})^{\prime}((a,b)^{\prime} - \bm{\mu}_{\Phi})}{2\sigma_{\Phi}^{2}}\right) \times \mathrm{exp}\left(-\frac{\bm{\eta}^{\prime}\textbf{K}^{-1}\bm{\eta}}{2}\right),\\
             \nonumber
 [\phi\vert \bm{\theta}_{-\phi},\bz, \{\sigma_{\ell,i}^{2}\}]  \propto & \hspace{4pt}               \mathrm{IG}\left(r/2 + \alpha_{\phi}, \omega_{\phi} + \bm{\eta}^{\prime}\bm{\Psi}^{\prime}\bm{\Phi}\bm{\Lambda}_{Q}\bm{\Phi}^{\prime}\bm{\Psi}\bm{\eta}/2\right)\\
              \nonumber
 [\sigma_{\epsilon,\ell,i}^{2}\vert \bm{\theta}_{-\epsilon,\ell,i},\bz, \{\sigma_{\ell,i}^{2}\}]  \propto & \hspace{4pt} \mathrm{IG}\left(1/2 + \alpha_{\epsilon}, \omega_{\epsilon} + (\mathrm{log}(\sigma_{\ell,i}^{2})-Y_{\ell,i})^{2}/2\right)\\
\label{fullcond}
 [\sigma_{\gamma}^{2}\vert \bm{\theta}_{-\gamma},\bz, \{\sigma_{\ell,i}^{2}\}]  \propto & \hspace{4pt} \mathrm{IG}\left(n_{1}/2 + \alpha_{\gamma}, \omega_{\gamma} + \sum_{i = 1}^{n_{1}}\xi_{i}^{2}/2\right),
  \end{align}
  \noindent
  which are the full-conditionals needed for the Metropolis-Hastings algorithm. Define
  \begin{equation*}
	  \alpha(\rho_{*},\rho_{k-1}^{*})\equiv \mathrm{min}\left\lbrace\frac{q_{\rho}(\rho_{k-1}^{*})[\rho_{*}\vert \bm{\theta}_{-\rho,k-1}^{*}, \bz]}{q_{\rho}(\rho_{*})[\rho_{k-1}^{*}\vert \bm{\theta}_{-\rho,k-1}^{*}, \bz]}\right\rbrace;\hspace{5pt}k = 1,...,K
  \end{equation*}
  \noindent
  where $\rho = \bm{\eta}, \bm{\beta}, (a,b)^{\prime},$ $\phi$, $\sigma_{\epsilon,\ell,i}^{2}$, and $\sigma_{\gamma}^{2}$, $\rho_{*}$ is a value generated from the generic target distribution $q_{\rho}(\cdot)$, and $\rho_{k}^{*}$ (and $\bm{\theta}_{-\rho,k}^{*}$) is the $k$-th MCMC replicate of $\rho$ (and $\bm{\theta}_{-\rho}$). We now provide step-by-step instructions for running this MCMC algorithm in the list below.}
  \begin{enumerate}
	  \item {Choose initial values for $\bm{\eta}$, $\bm{\beta}$, $\{\xi_{i}\}$, $(a,b)^{\prime}$, $\phi$, $\{\sigma_{\epsilon,\ell,i}^{2}\}$, and $\sigma_{\gamma}^{2}$ and denote them with $\bm{\eta}_{0}^{*}$, $\bm{\beta}_{0}^{*}$, $\{\xi_{i}\}_{0}^{*}$, $(a_{0}^{*},b_{0}^{*})^{\prime}$, $\phi_{0}^{*}$, $\{\sigma_{\epsilon,\ell,i}^{2}\}_{0}^{*}$, and $(\sigma_{\gamma}^{2})_{0}^{*}$ respectively. Let $k$ = 0.}
	  \item {Let $k = k+1$.}
	  \item {Update the components of $\bm{\eta}$.}
	  	\begin{enumerate}
		  	\item {Generate a proposed value of $\bm{\eta}_{*}$ from target distribution $q_{\eta}(\bm{\eta})$.}
		  	\item {Let $\bm{\eta}_{k}^{*} = \bm{\eta}_{*}$ with probability $\mathrm{min}\left\lbrace 1,\alpha(\bm{\eta}_{*},\bm{\eta}_{k-1}^{*})\right\rbrace$ and let $\bm{\eta}_{k}^{*} = \bm{\eta}_{k-1}^{*}$ with probability $1-\mathrm{min}\left\lbrace 1,\alpha(\bm{\eta}_{*},\bm{\eta}_{k-1}^{*})\right\rbrace$.}
	  	\end{enumerate}
	  \item {Update the components of $\bm{\beta}$.}
	  	\begin{enumerate}
		  	\item {Generate a proposed value of $\bm{\beta}_{*}$ from target distribution $q_{\beta}(\bm{\beta})$.}
		  	\item {Let $\bm{\beta}_{k}^{*} = \bm{\beta}_{*}$ with probability $\mathrm{min}\left\lbrace 1,\alpha(\bm{\beta}_{*},\bm{\beta}_{k-1}^{*})\right\rbrace$ and let $\bm{\beta}_{k}^{*} = \bm{\beta}_{k-1}^{*}$ with probability $1-\mathrm{min}\left\lbrace 1,\alpha(\bm{\beta}_{*},\bm{\beta}_{k-1}^{*})\right\rbrace$.}
	  	\end{enumerate}
	  \item {Update the components of $\{\xi_{i}\}$.}
	  	\begin{enumerate}
		  	\item {Generate a proposed value of $\{\xi_{i}\}_{*}$ from target distribution $q_{\beta}(\{\xi_{i}\})$.}
		  	\item {Let $\{\xi_{i}\}_{k}^{*} = \{\xi_{i}\}_{*}$ with probability $\mathrm{min}\left\lbrace 1,\alpha(\{\xi_{i}\}_{*},\{\xi_{i}\}_{k-1}^{*})\right\rbrace$ and let $\{\xi_{i}\}_{k}^{*} = \{\xi_{i}\}_{k-1}^{*}$ with probability $1-\mathrm{min}\left\lbrace 1,\alpha(\{\xi_{i}\}_{*},\{\xi_{i}\}_{k-1}^{*})\right\rbrace$.}
	  	\end{enumerate}
	  \item {Update the components of $(a,b)^{\prime}$.}
	  	\begin{enumerate}
		  	\item {Generate a proposed value of $(a_{*},b_{*})^{\prime}$ from target distribution $q_{(a,b)}((a,b)^{\prime})$.}
		  	\item {Let $(a_{k}^{*},b_{k}^{*})^{\prime} = (a_{*},b_{*})^{\prime}$ with probability $\mathrm{min}\left\lbrace 1,\alpha((a_{*},b_{*})^{\prime},(a_{k-1}^{*},b_{k-1}^{*})^{\prime})\right\rbrace$ and let $(a_{k}^{*},b_{k}^{*})^{\prime} = (a_{k-1}^{*},b_{k-1}^{*})^{\prime}$ with probability $1-\mathrm{min}\left\lbrace 1,\alpha((a_{*},b_{*})^{\prime},(a_{k-1}^{*},b_{k-1}^{*})^{\prime})\right\rbrace$.}
	  	\end{enumerate}
	  \item {Update $\phi$, $\{\sigma_{\epsilon,\ell,i}^{2}\}$, and $\sigma_{\gamma}^{2}$.}
	  	\begin{enumerate}
		  	\item {Let $\phi_{k}^{*}$ equal to a value generated from the full conditional of $\phi$ in (\ref{fullcond}).}
		  	\item {Generate $\sigma_{\epsilon,\ell,i}^{2}$ from it's full conditional in (\ref{fullcond}) for each $\ell = 1,...,L$ and each $i = 1,...,n_{\ell}$; denote these values as $(\sigma_{\epsilon,\ell,i}^{2})_{k}^{*}$. Then set $\{\sigma_{\epsilon,\ell,i}^{2}\}_{k}^{*} = \{(\sigma_{\epsilon,\ell,i}^{2})_{k}^{*}: \ell = 1,...,L, i = 1,...,n_{\ell}\}$}.
		  	\item {Let $(\sigma_{\gamma}^{2})_{k}^{*}$ equal to a value generated from the full conditional of $\sigma_{\gamma}^{2}$ in (\ref{fullcond}).} 
	  	\end{enumerate}
	  \item {Repeat steps 2-7 until convergence.}
\end{enumerate}

\newpage
\singlespacing

\begin{figure}[H]
  \begin{center}
  \begin{tabular}{c}
   \includegraphics[width=8cm,height=8cm]{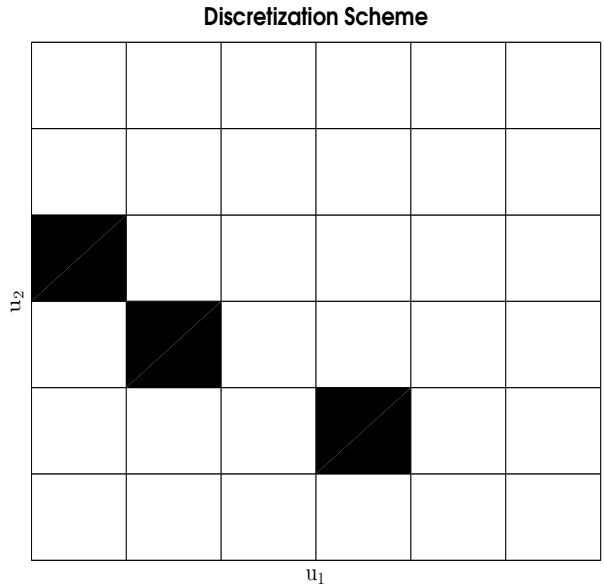}
  \end{tabular}
  \caption{We show a discretization of the spatial domain $D=\{\bu: \bu = (u_{1},u_{2})^{\prime}, u_{1}\in [1,12],u_{2}\in [1,12]\}$. Three grid-cells were randomly chosen to represent areas of $D$ that have a high number of psuedo-households (called hot-spots), which are colored back; the remaining cells are colored white.}\label{fig2}
  \end{center}
  \end{figure}
  
  \newpage
    \begin{figure}[H]
    \begin{center}
    \begin{tabular}{cc}
    \includegraphics[width=5.25cm,height=5.25cm]{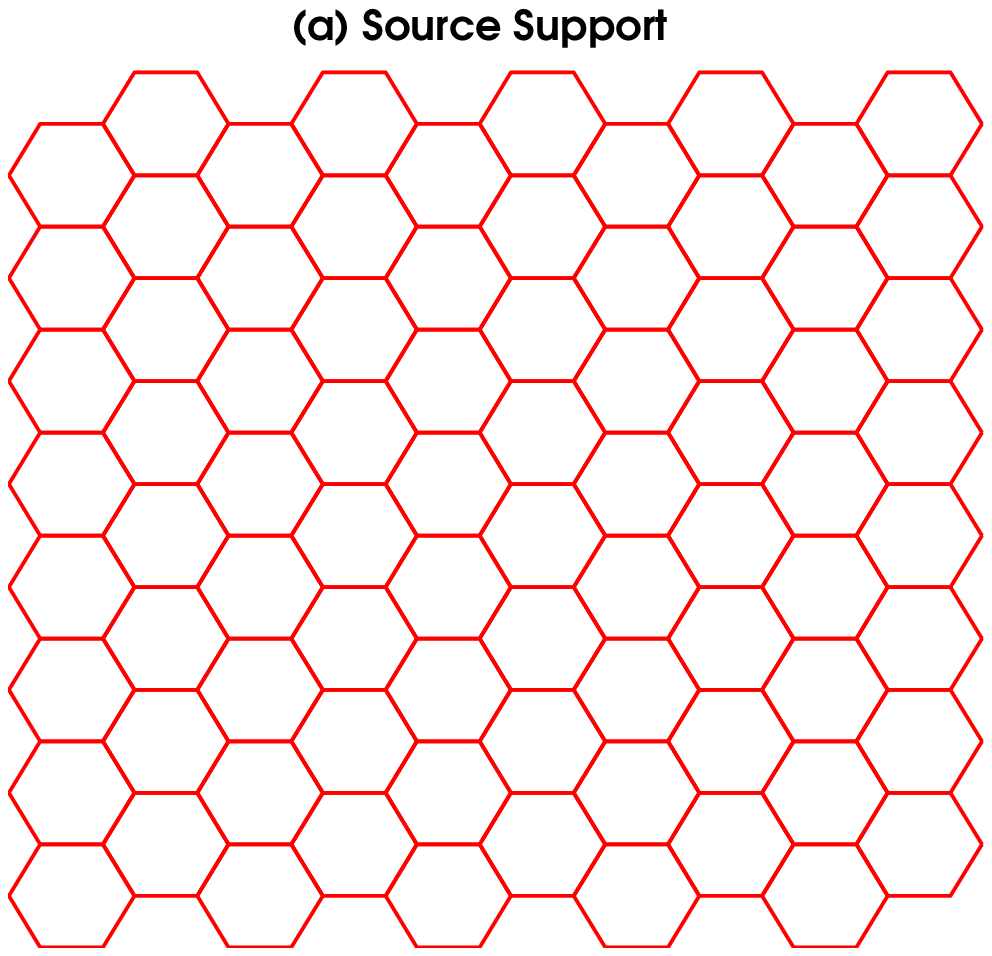}& \includegraphics[width=5.25cm,height=5.25cm]{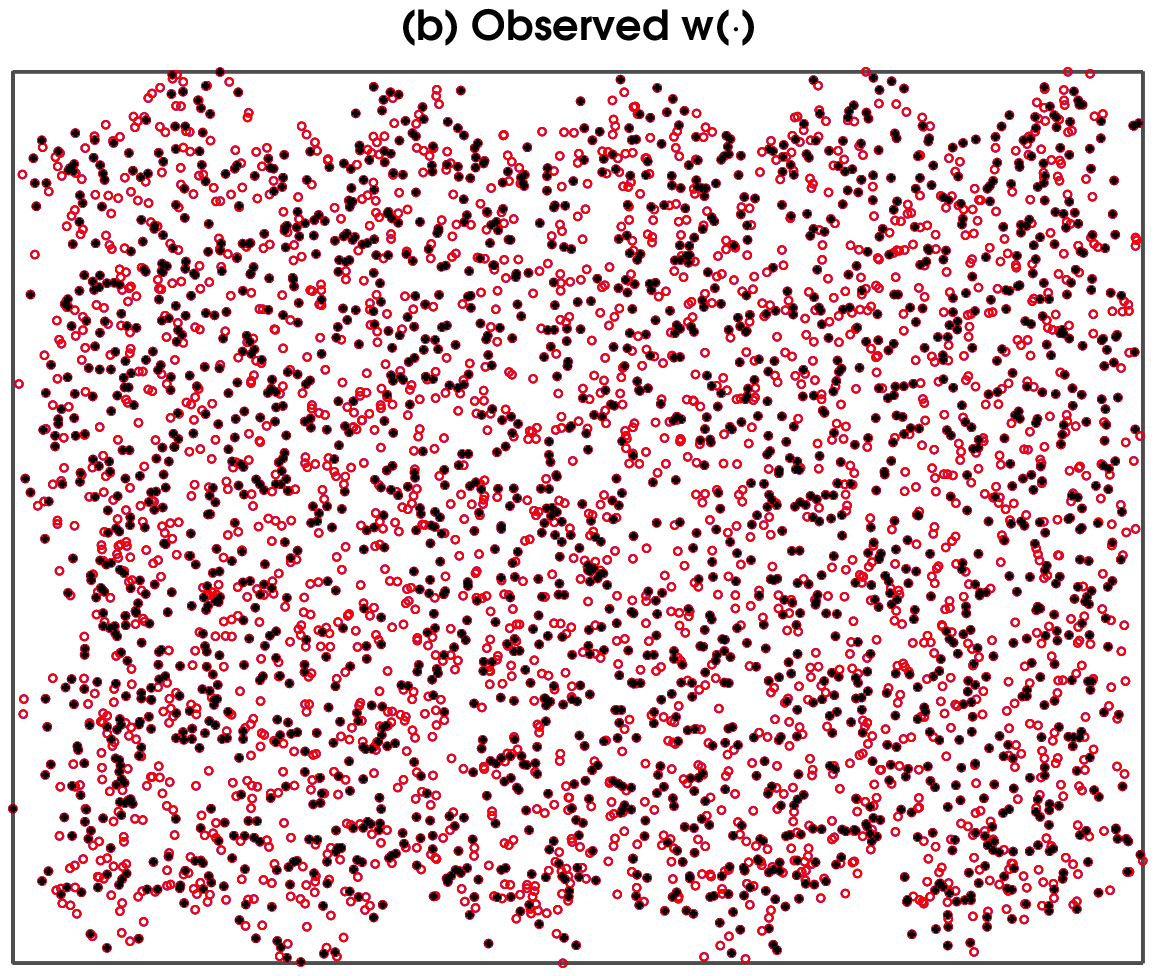}\\
    \includegraphics[width=5.75cm,height=5.75cm]{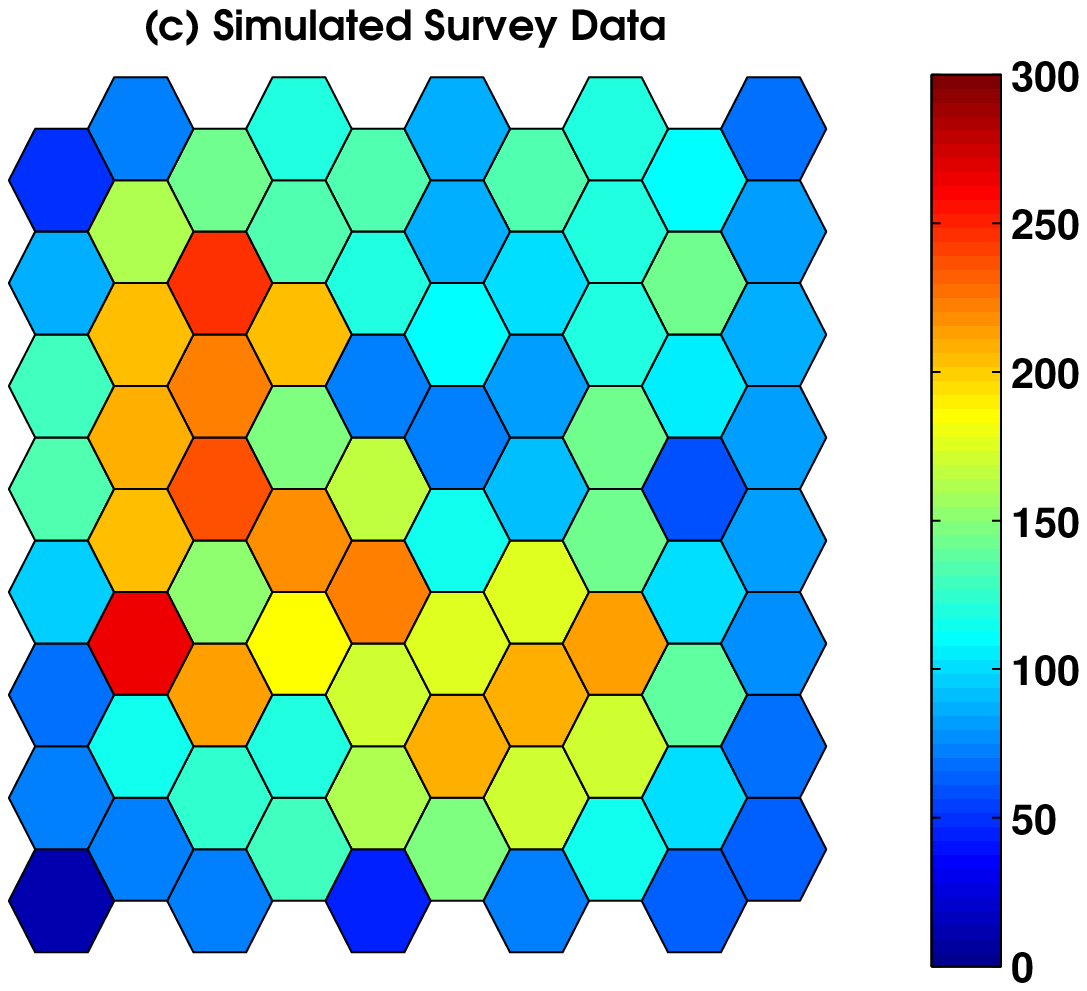}&  \includegraphics[width=5.75cm,height=5.75cm]{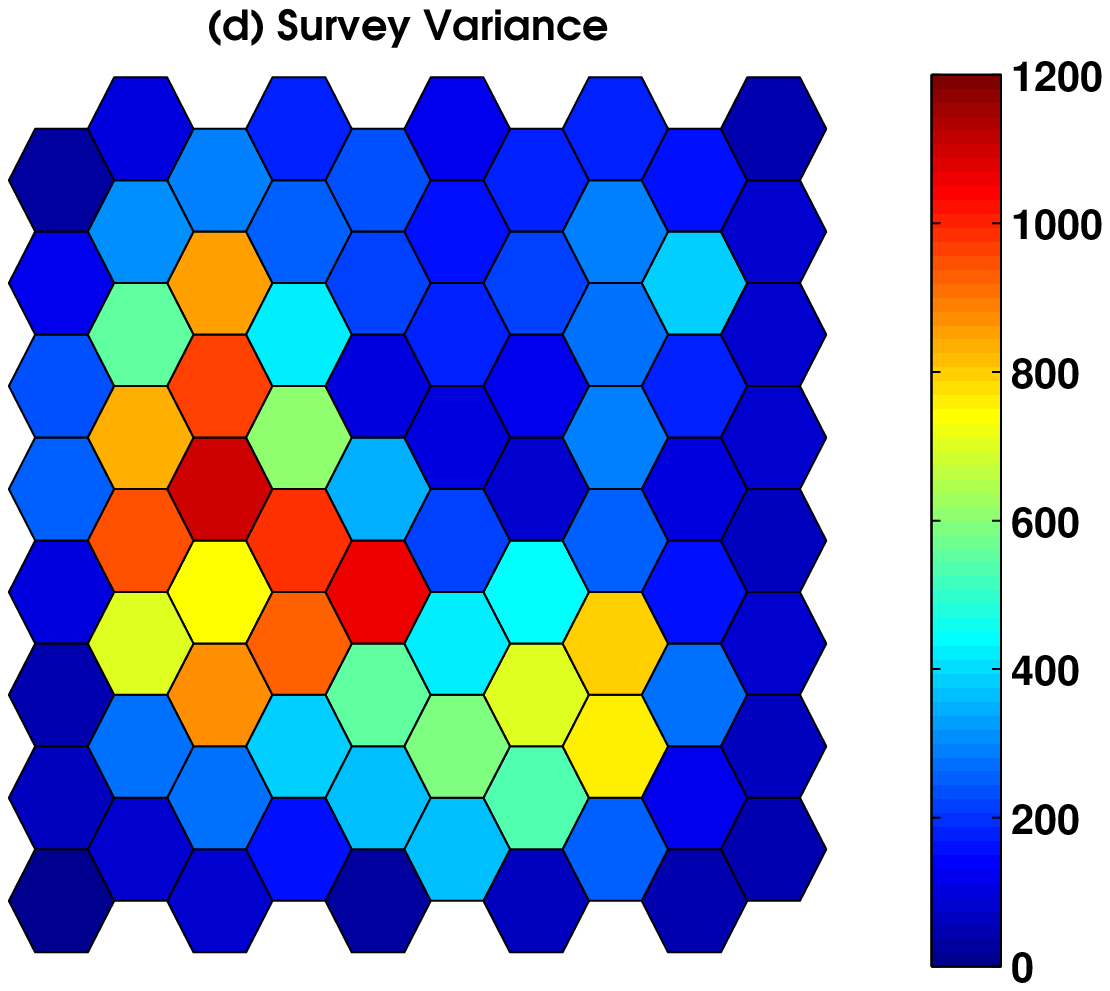}\\
    \includegraphics[width=5.25cm,height=5.25cm]{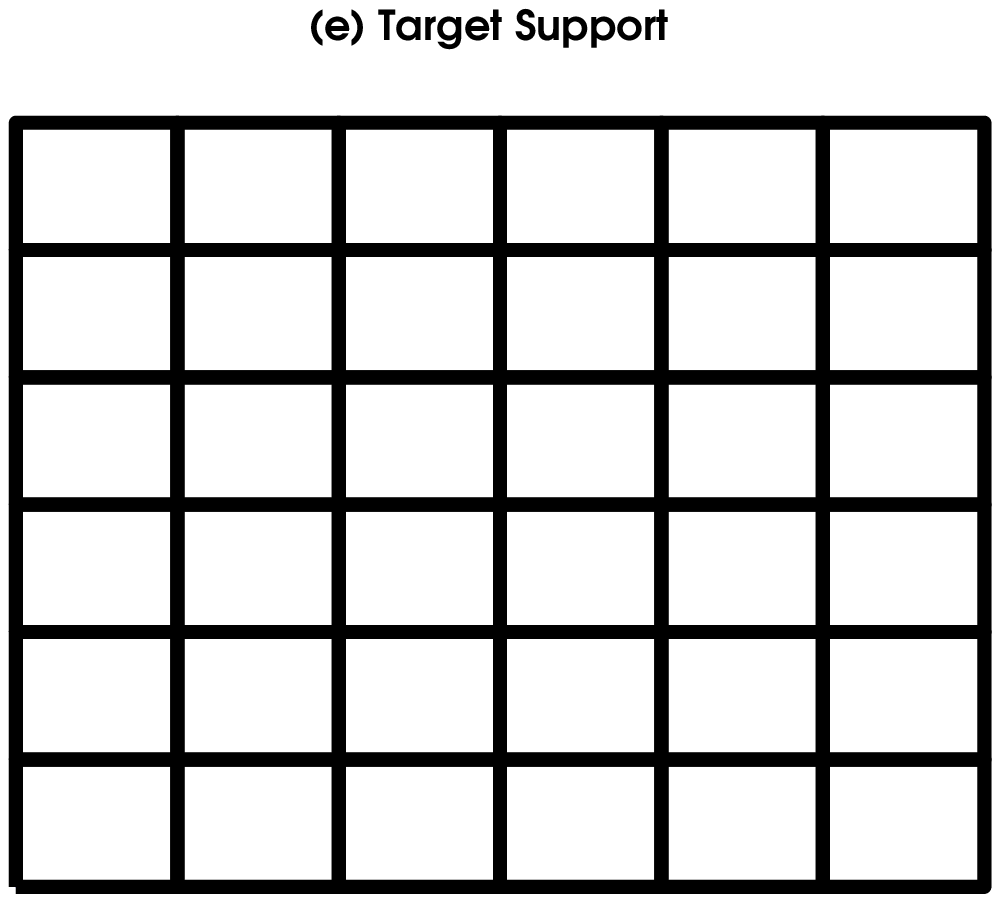} & \includegraphics[width=5.25cm,height=5.25cm]{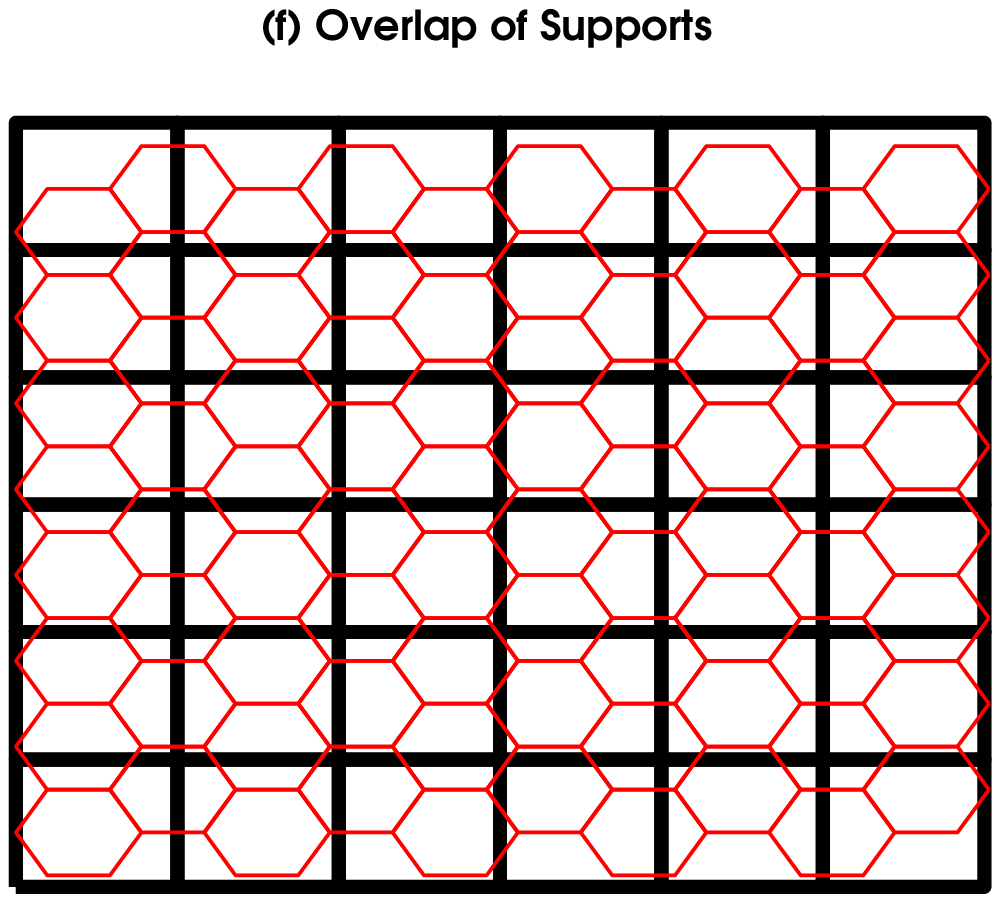}
    \end{tabular}
    \caption{Each polygon in (a) represents an areal unit on the source support (i.e., $A_{\ell,1},...,A_{\ell,90}$). In (b) we sample 50 values of $w(\cdot)$ within each strata to obtain $\{\bs_{i,j}:i = 1,...,90,j = 1,...,50\}$ and $\{w(\bs_{i,j}):i = 1,...,90,j = 1,...,50\}$; in (b) we indicate where $w(\bs) = 0$ and $w(\bs) = 1$ with the colors red and black, respectively. In (c) and (d), we display the sets $\{Z(A_{\ell,i})\}$ and $\{\sigma_{\ell,i}^{2}\}$ computed using the realized values of $\{w(\bs_{i}): i = 1,...,n\}$ in (b), Equations (\ref{survey:weights}) and (\ref{variance:survey:weights}), and the strata in (a). Each polygon in (e) represents an areal unit on the target support (i.e., $A_{\ell+1,1},...,A_{\ell+1,36}$). In (f), we have the overlap between the source and target supports.}
    \end{center}
    \end{figure}

  \newpage
    \begin{figure}[H]
    \begin{center}
    \begin{tabular}{ccc}
    \includegraphics[width=4.2cm,height=4.5cm]{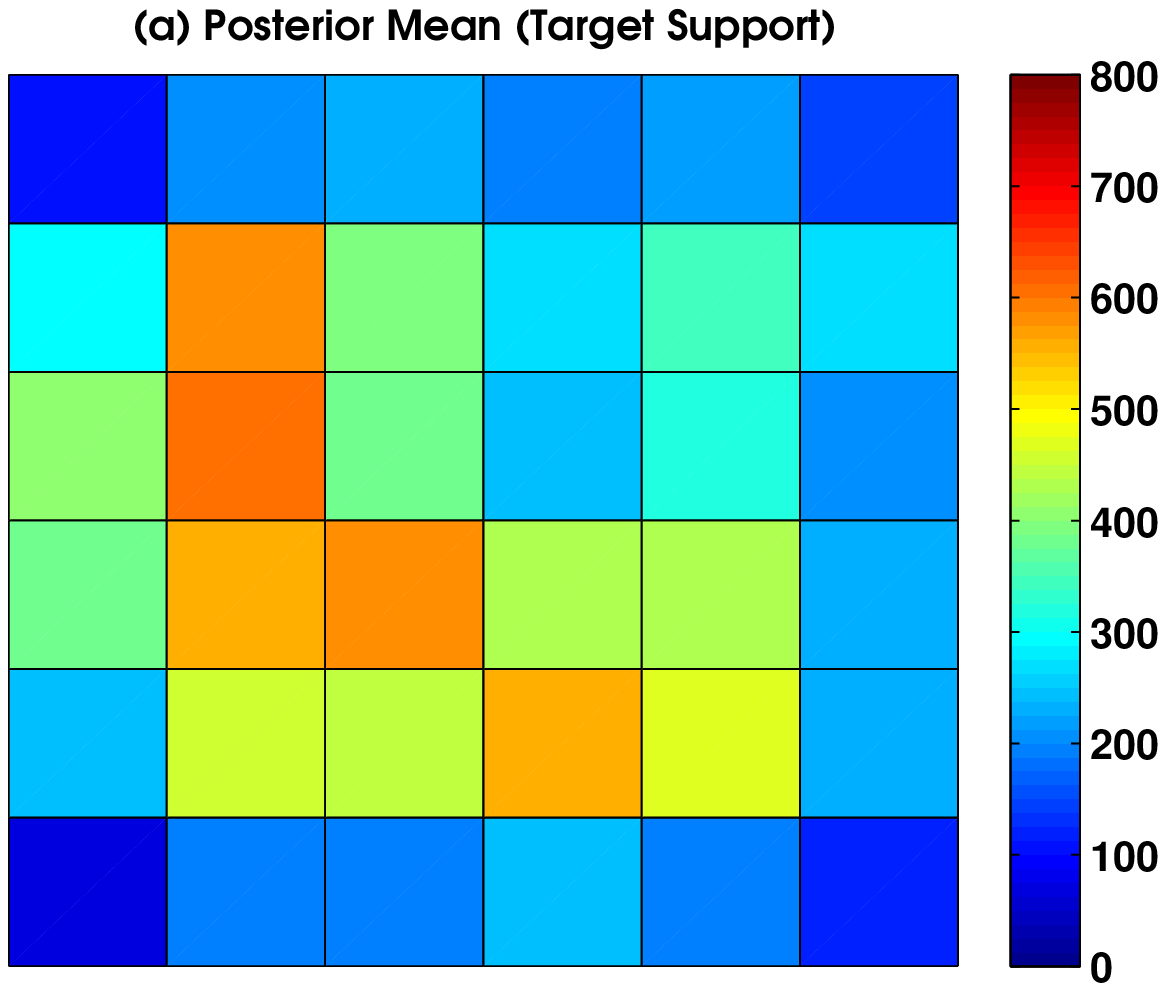} & \hspace{-2pt}\includegraphics[width=4.2cm,height=4.5cm]{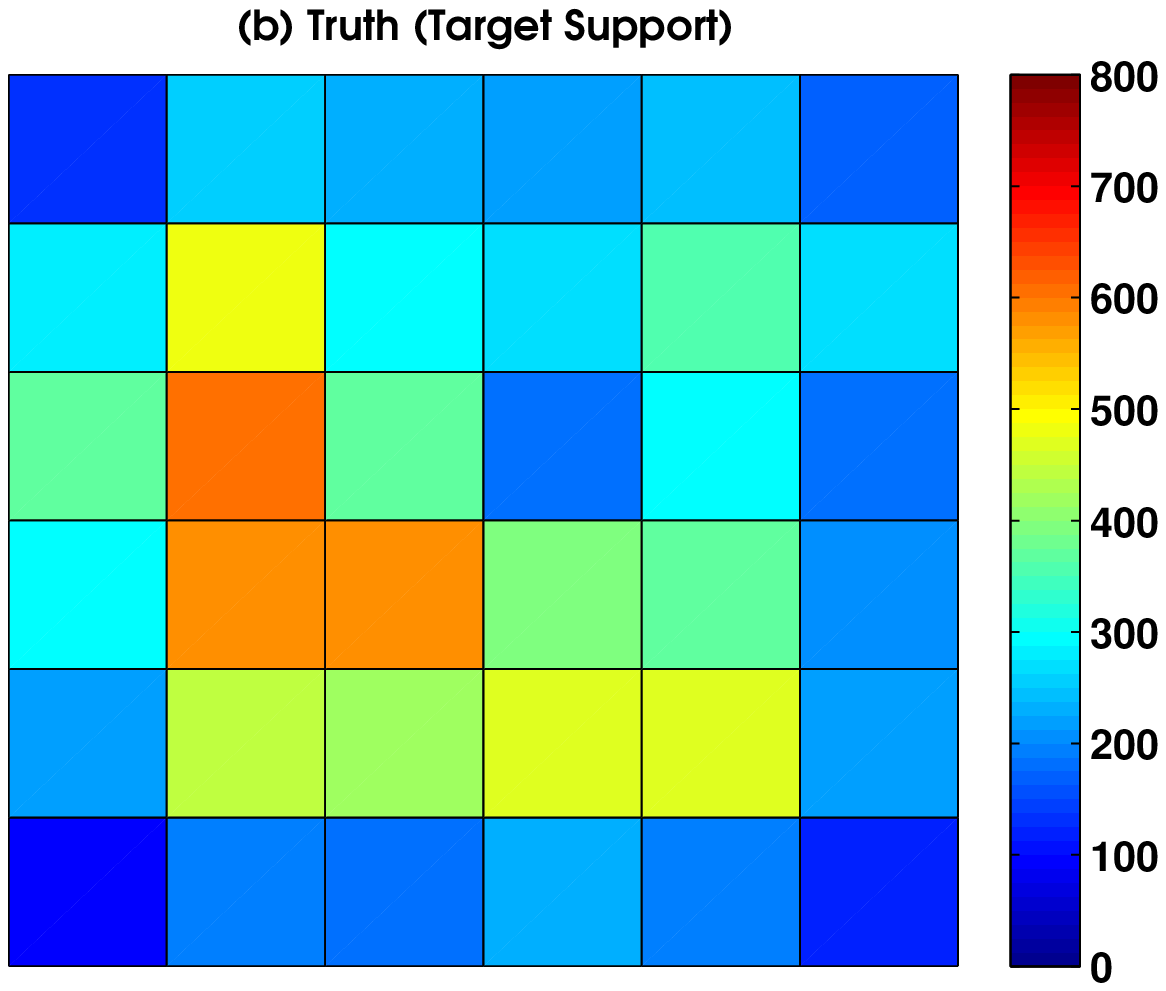} & \includegraphics[width=4.2cm,height=4.5cm]{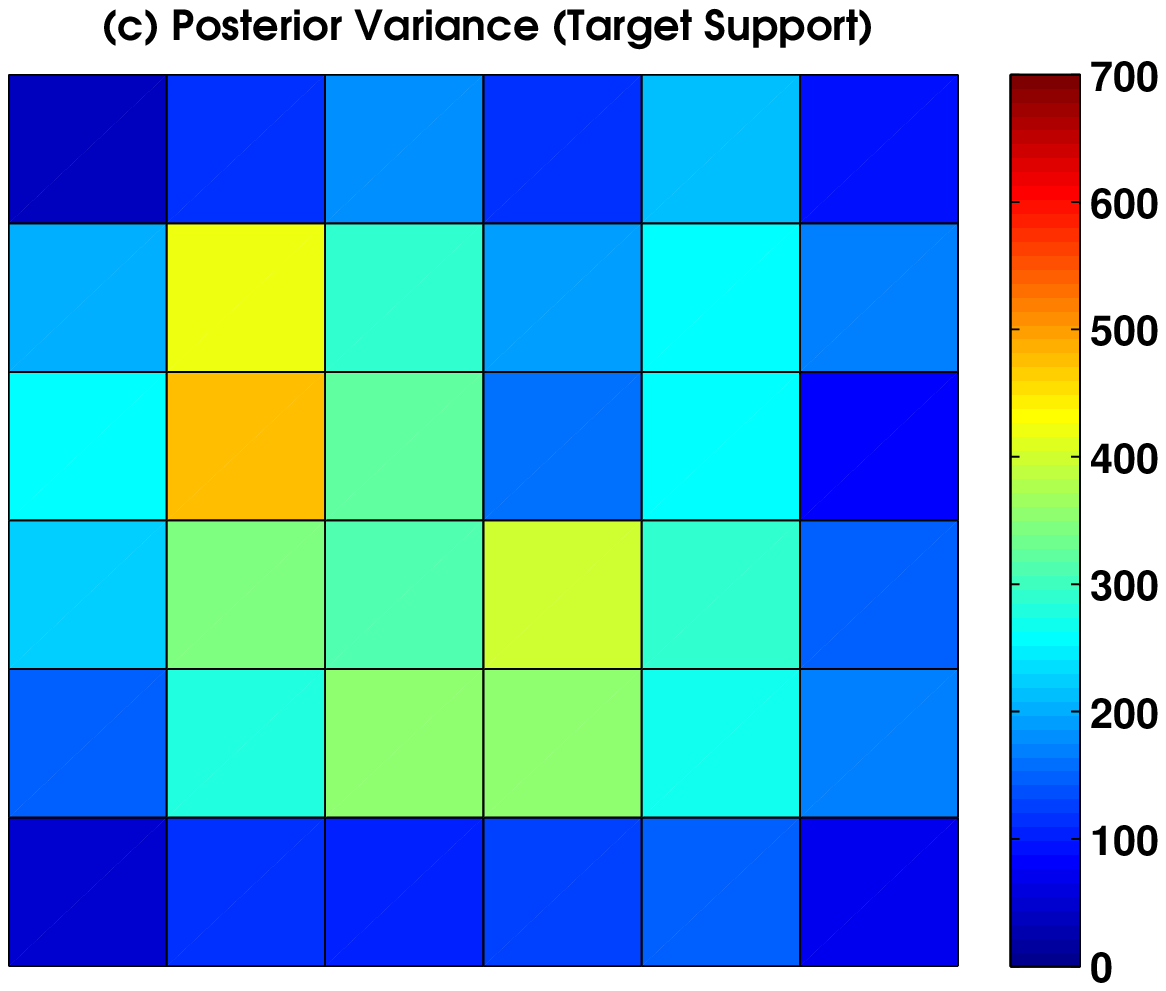} \\
    \includegraphics[width=5cm,height=4.5cm]{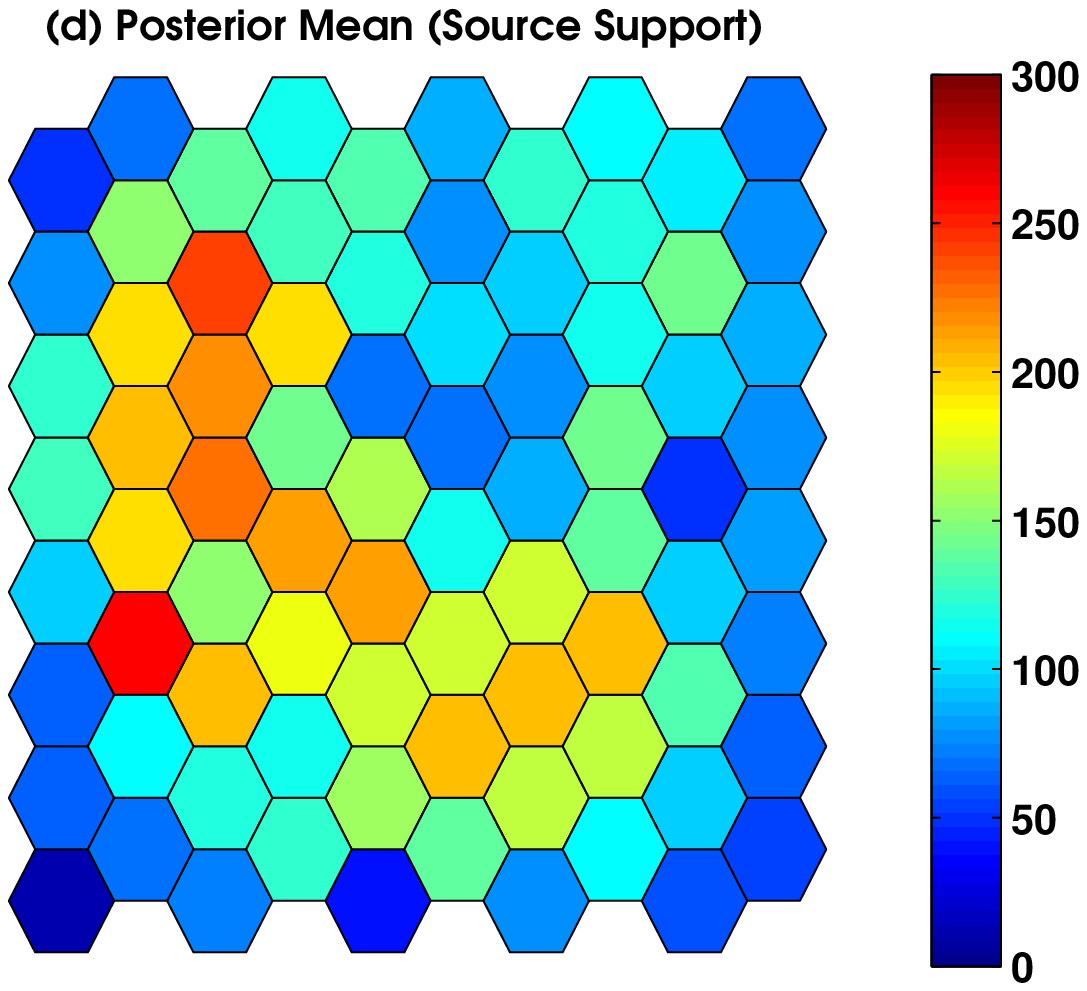} & \hspace{-2pt}\includegraphics[width=5.2cm,height=4.5cm]{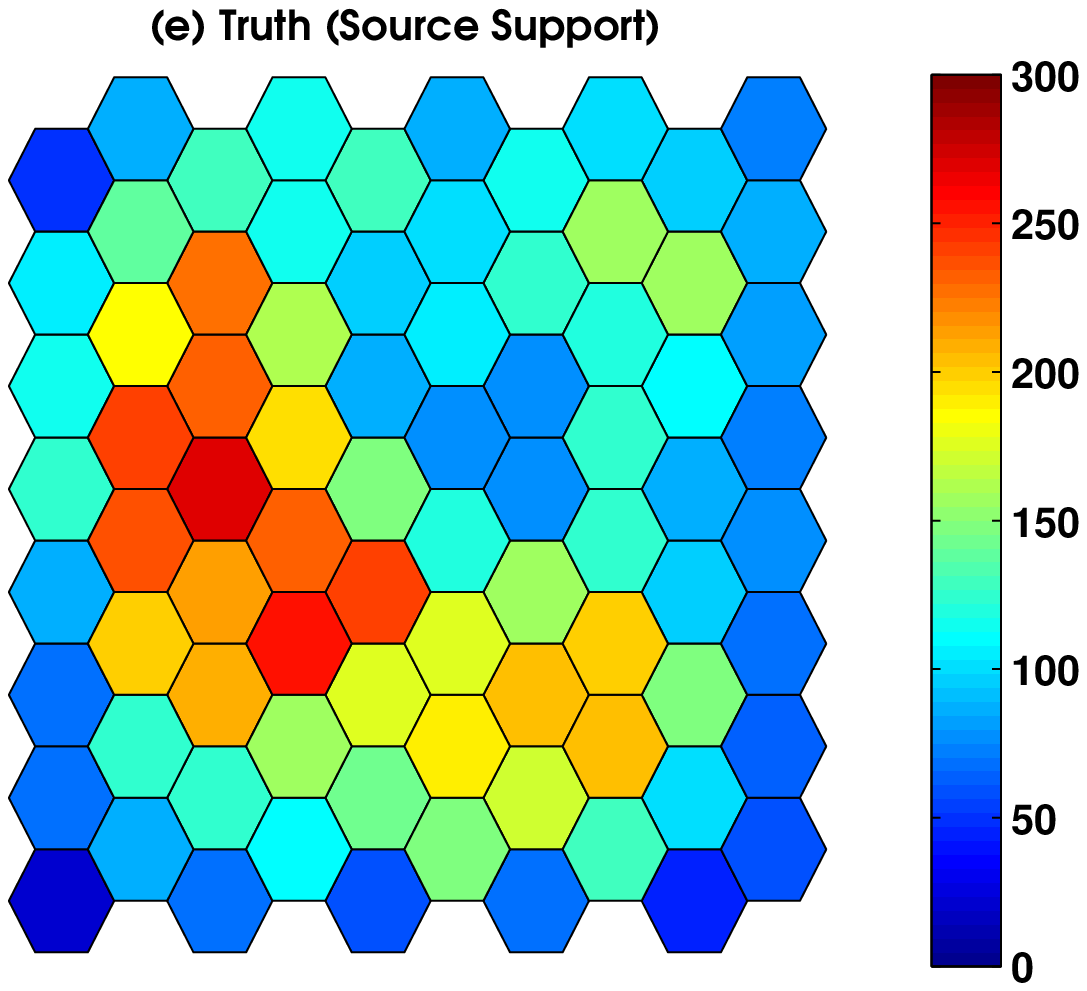} &\hspace{-2pt} \includegraphics[width=5.2cm,height=4.5cm]{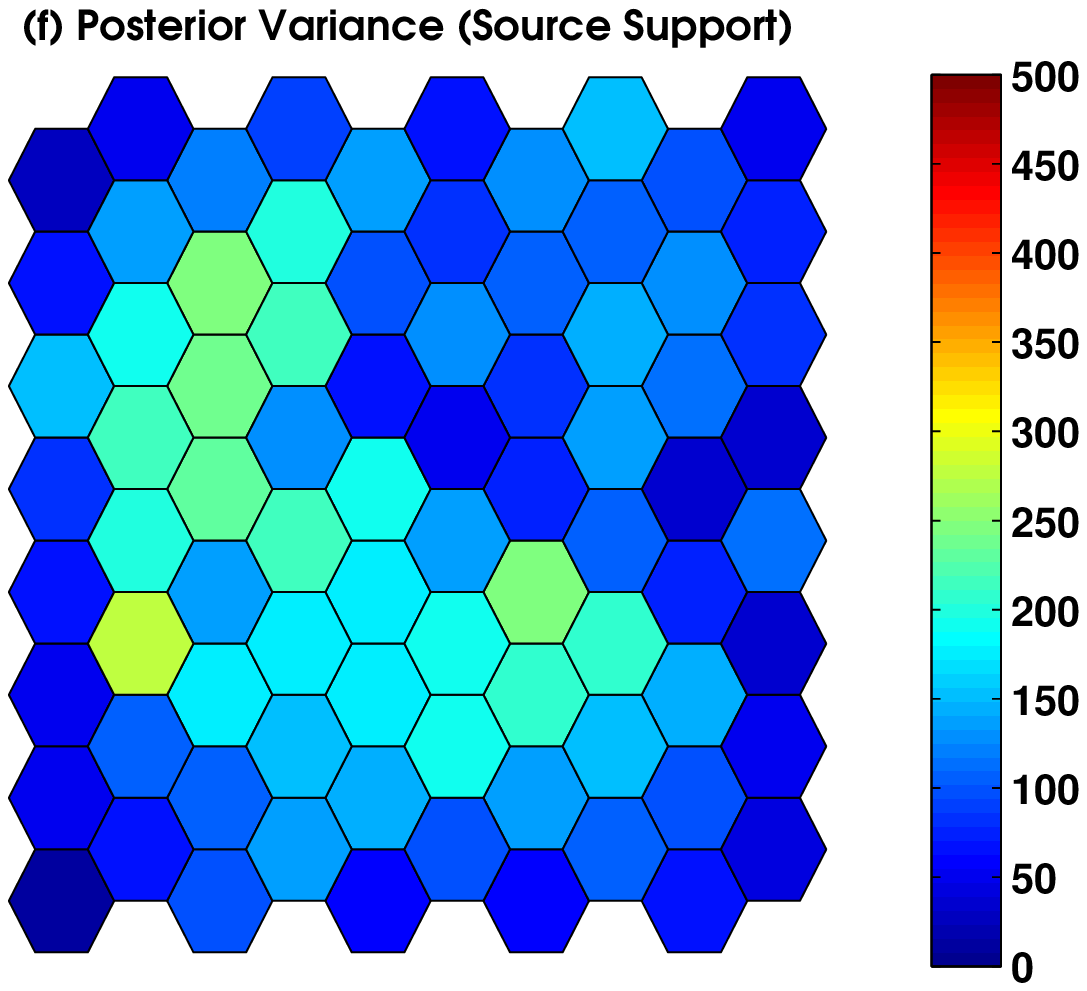}\\
    \includegraphics[width=4.2cm,height=4.5cm]{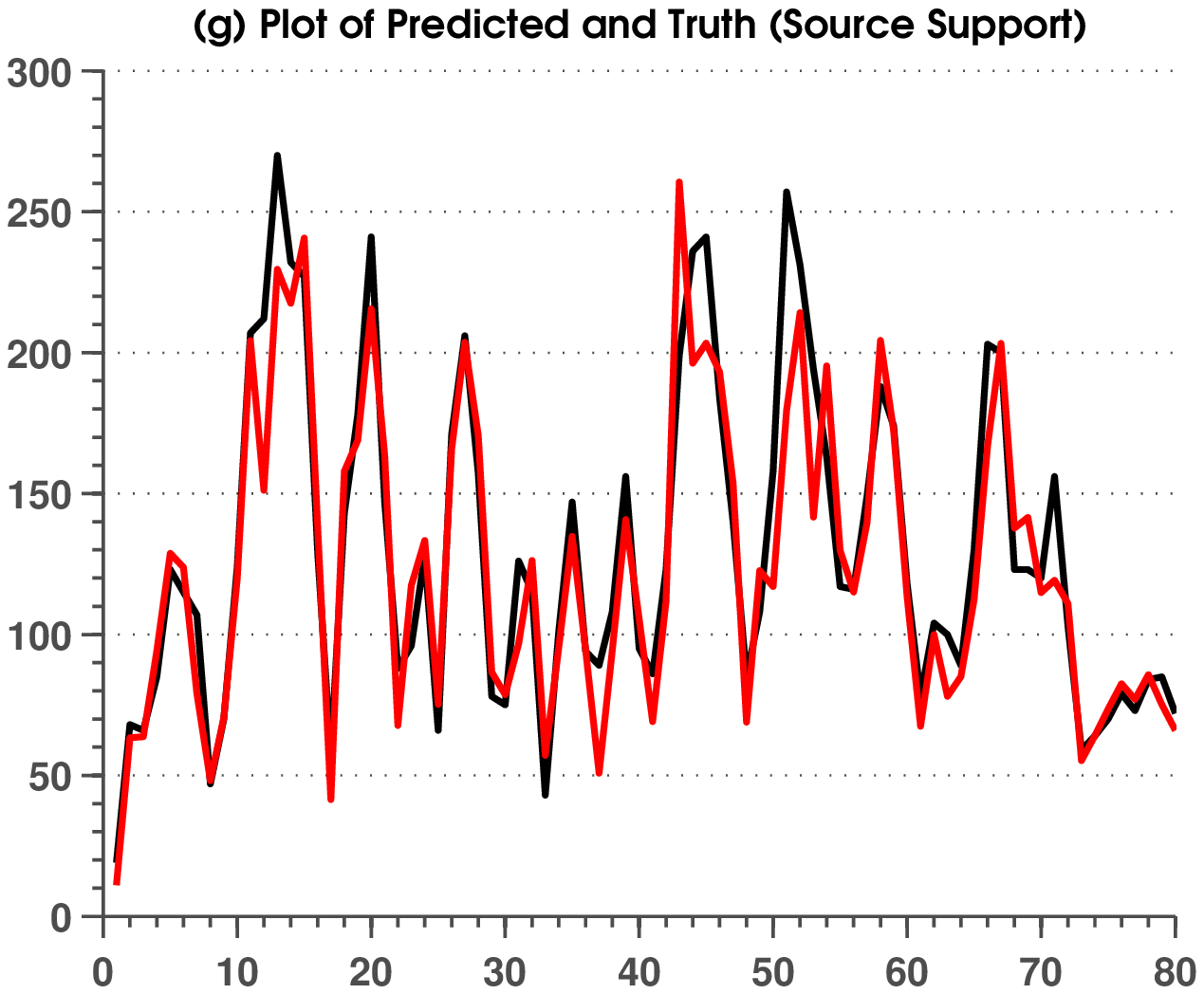}&\includegraphics[width=4.2cm,height=4.5cm]{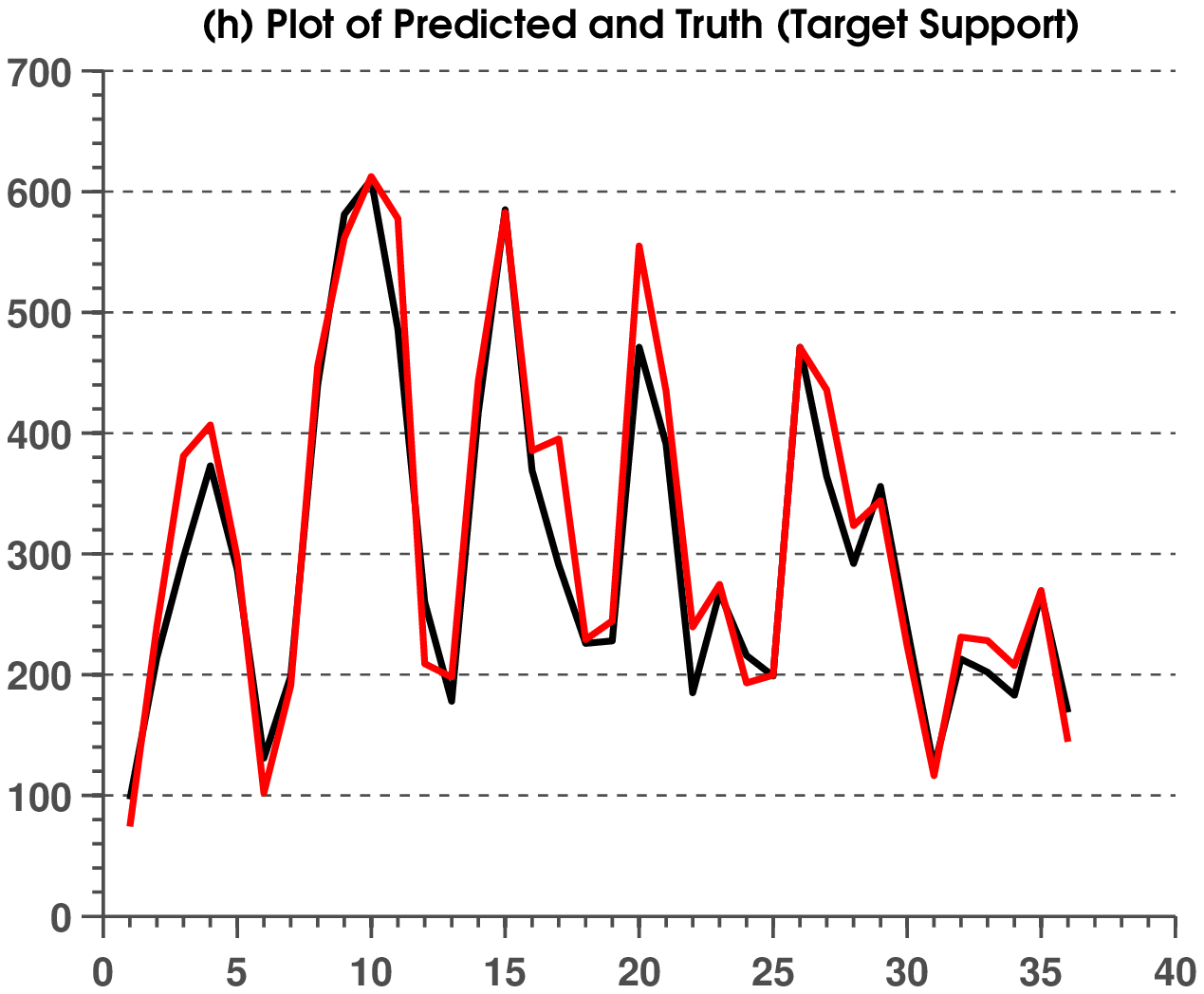} & \hfill
    \end{tabular}
    \caption{In (a), we present the posterior mean of $\{\mu(A_{\ell,i})\}$, and in (c) we present the posterior variance of $\{\mu(A_{\ell,i})\}$. In (d), we present the posterior mean of $\{\mu(B_{m})\}$, and in (f) we present the posterior variance of $\{\mu(B_{m})\}$. The respective true fields, known from the simulation, are given in (b) and (e). Note that the color scale between the two rows are different. In (g) we plot $\{\mu(A_{\ell,i})\}$ (black line) and the corresponding posterior mean (red line). In the (h) we plot $\{\mu(B_{m})\}$ (black line) and the corresponding posterior mean (red line).}\label{fig5}
    \end{center}
    \end{figure}

    \newpage
        \begin{figure}[H]
        \begin{center}
        \begin{tabular}{cc}
                \includegraphics[width=7cm,height=7cm]{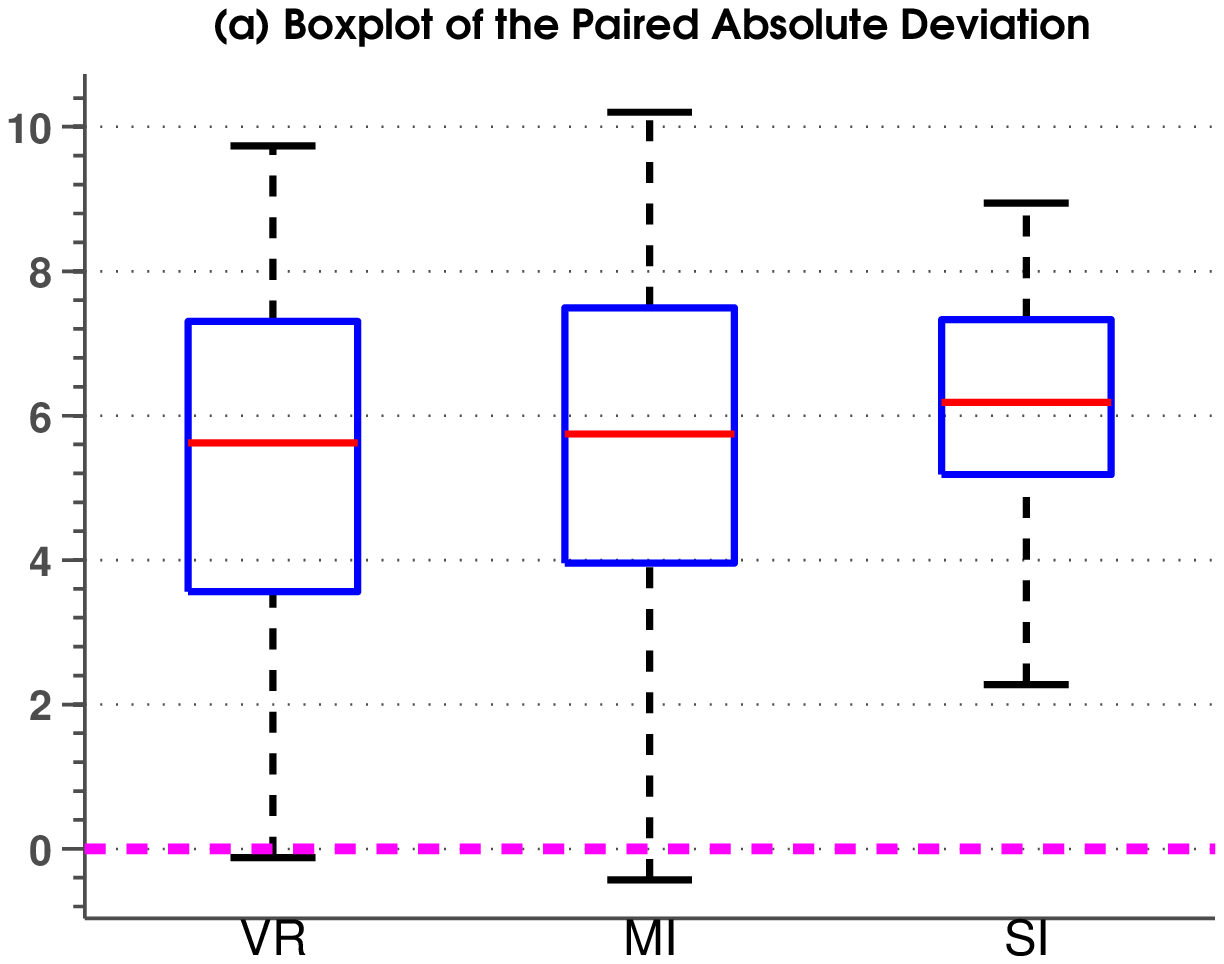} & \includegraphics[width=7cm,height=7cm]{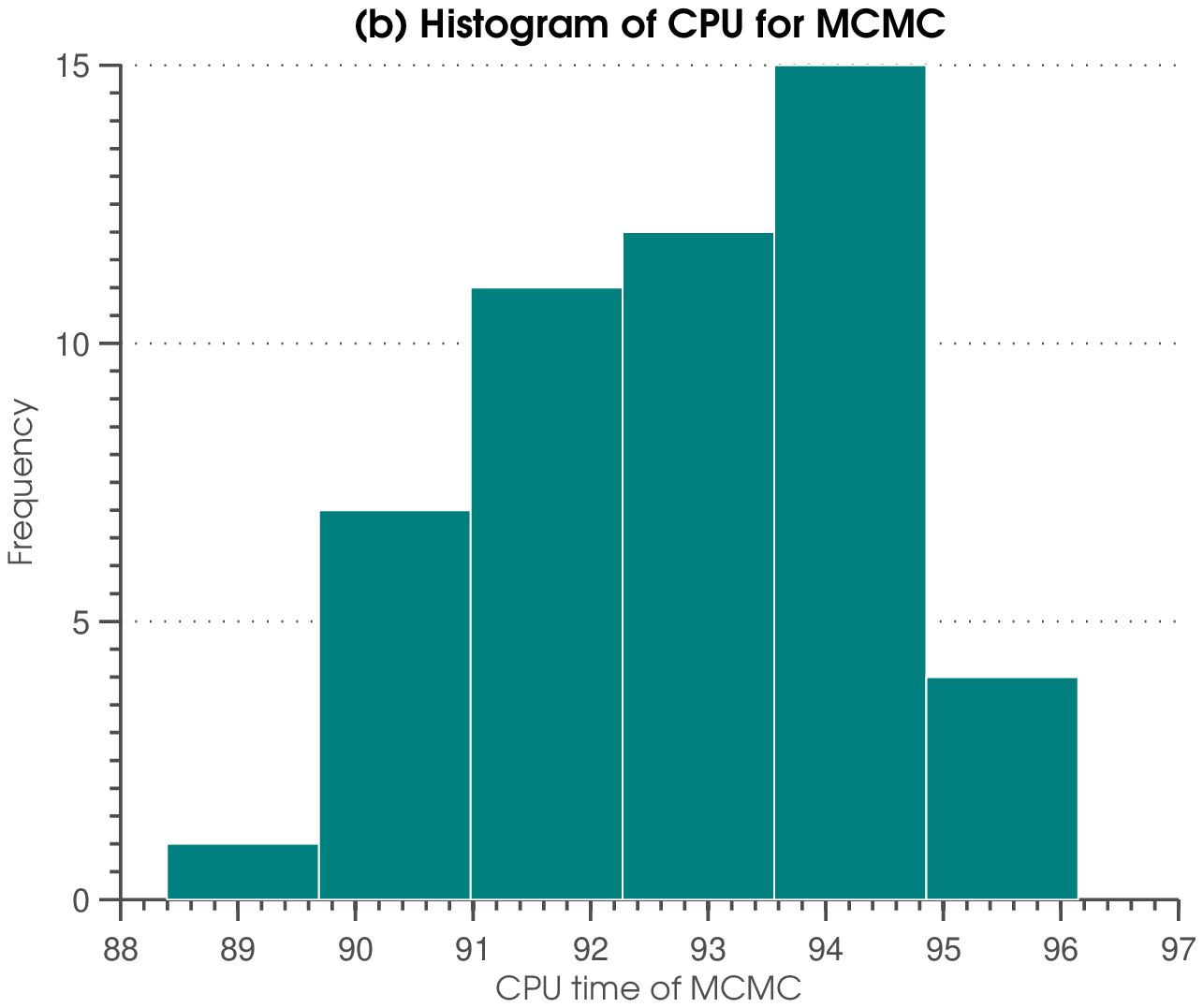}\\
        \includegraphics[width=7cm,height=7cm]{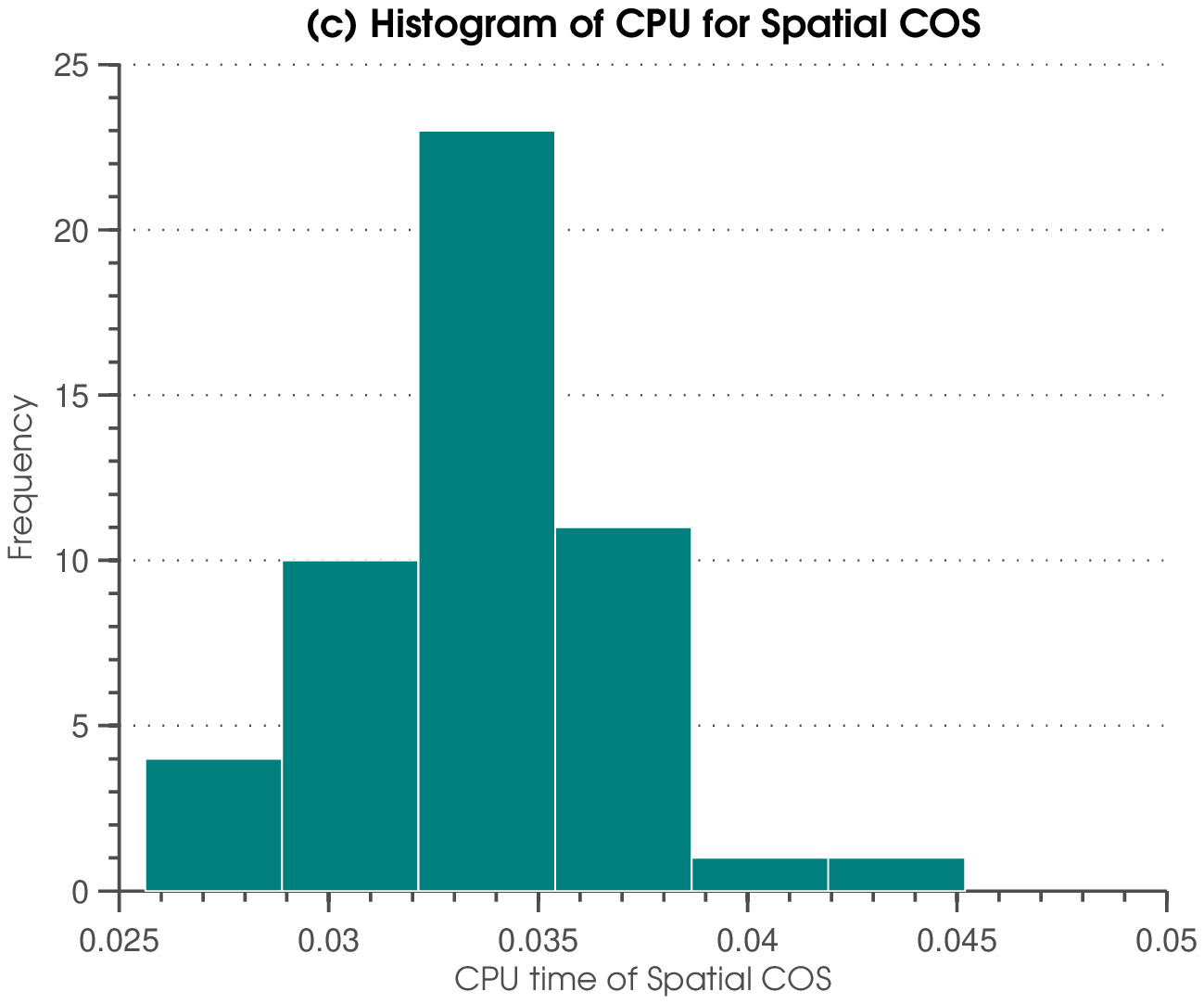}& \hfill
        \end{tabular}
        \caption{In (a), we present boxplots of PAD in (\ref{cht4.response}), by model MI (Moran's I), VR (variance removed), and SI (simple areal interpolation), and over 50 replicates of $\bz$. The steps used to generate each of the 50 $\bz$s are outlined in Section 3. The dotted magenta line references a PAD of 0; PAD above (below) this line indicates that model CS (MD) yields better results than MD (CS). In (b) and (c), we present histograms of CPU times {(in seconds)} over the 50 replicates of $\bz$. The steps used to generate each $\bz$ are outlined in Section 3. In (b), we provide a histogram of the CPU time (in seconds) required for MCMC computation of the BHM in (11). In (c), we provide a histogram of the CPU time (in seconds) required for spatial COS using Equation (\ref{COS2}).}\label{fig2}
        \end{center}
        \end{figure}
    
\newpage
      \begin{figure}[H]
      \begin{center}
      \begin{tabular}{cc}
      \includegraphics[width=7.5cm,height=8cm]{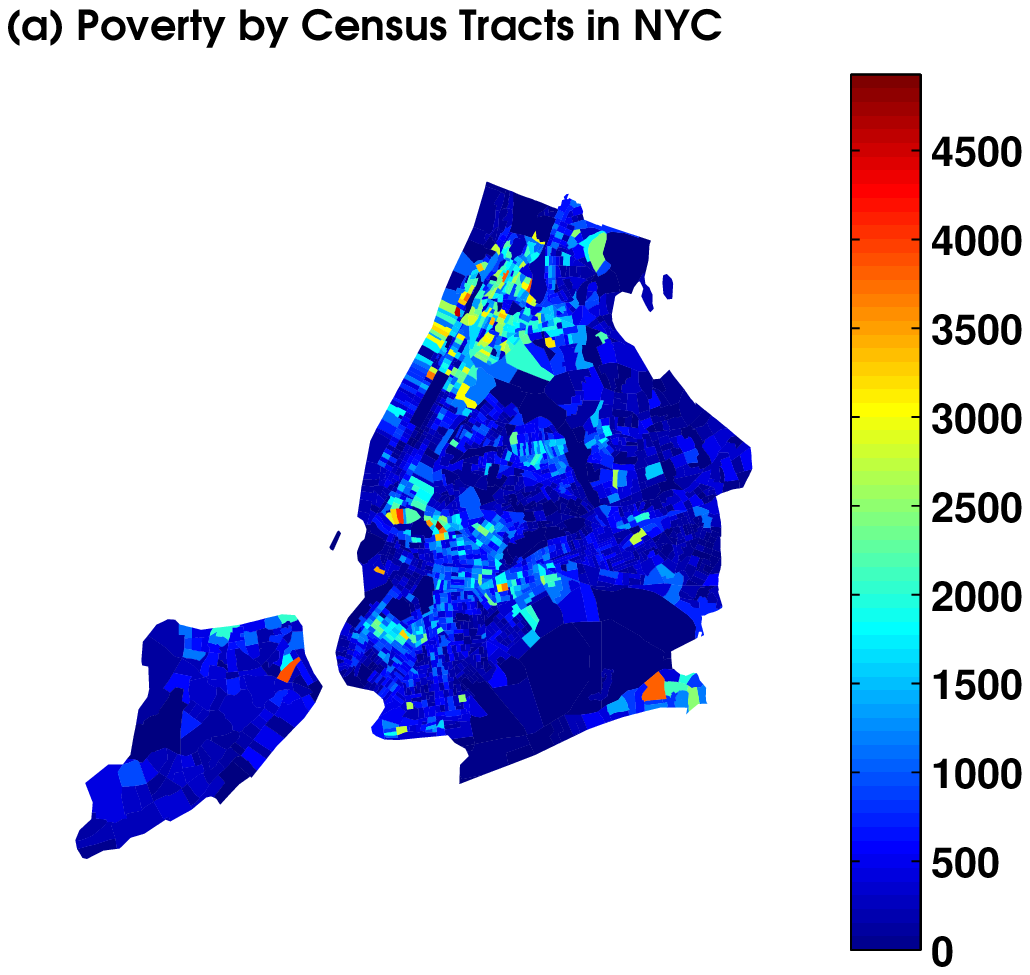}&
      \includegraphics[width=7.5cm,height=8cm]{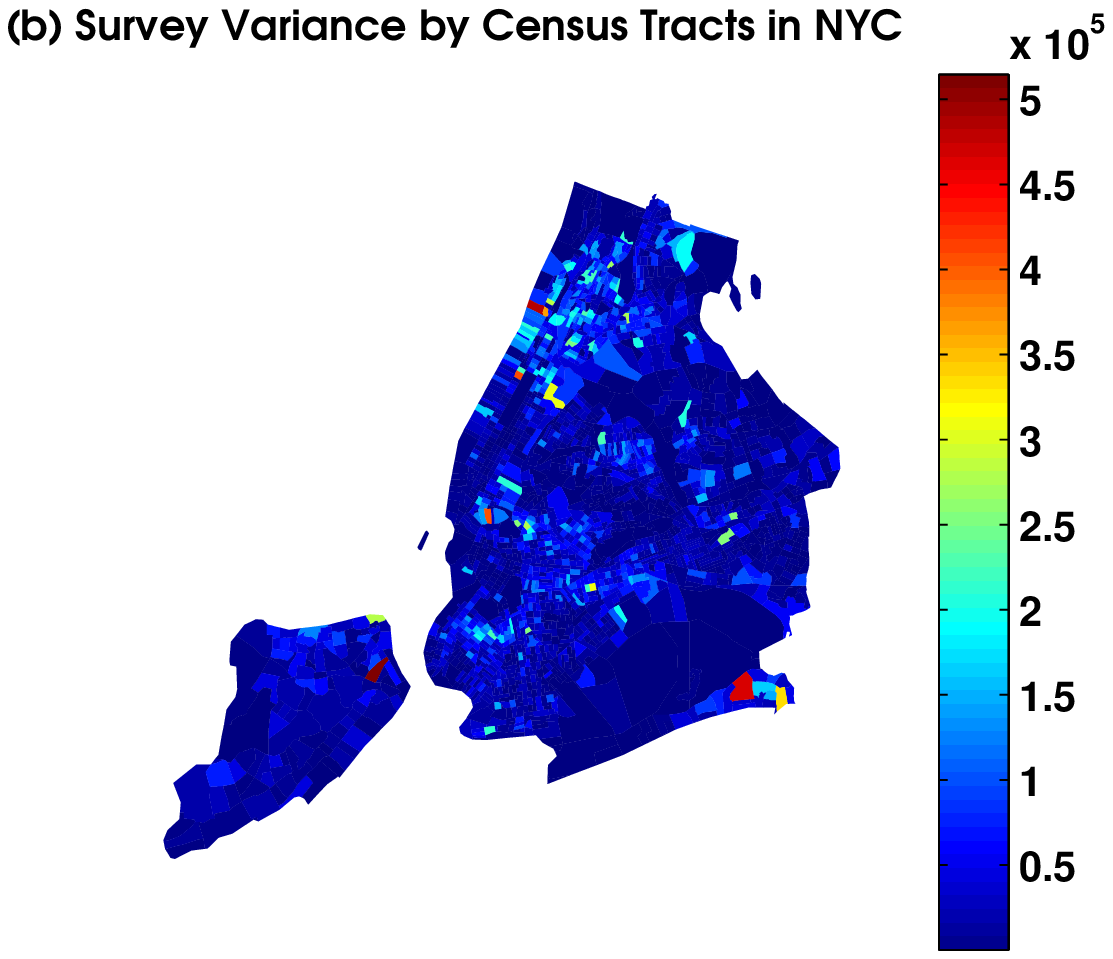}\\
            \includegraphics[width=7cm,height=7cm]{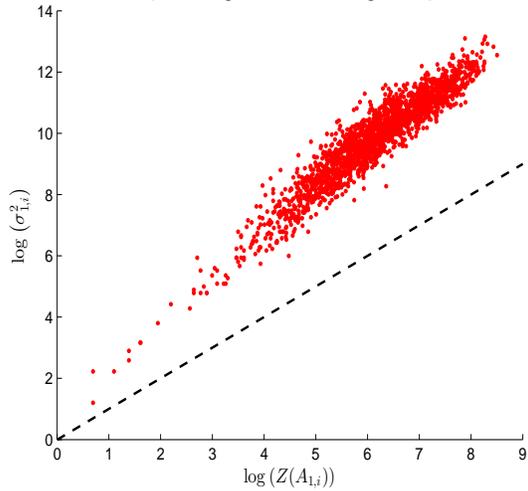}&\hfill
      \end{tabular}
      \end{center}
      \caption{The 2012 ACS 5-year period estimates of the population below the poverty threshold (left column), and the corresponding survey-based variance estimates (right column). These values are recorded over New York City's Census Tracts in (a) and (b). In (c), we provide a scatterplot of the log count data and log survey variance estimate. The dashed black line in (c) represents the line $y=x$.}\label{fig:6}
      \end{figure}

    \newpage
      \begin{figure}[H]
      \begin{center}
      \begin{tabular}{cc}
    \includegraphics[width=8cm,height=6cm]{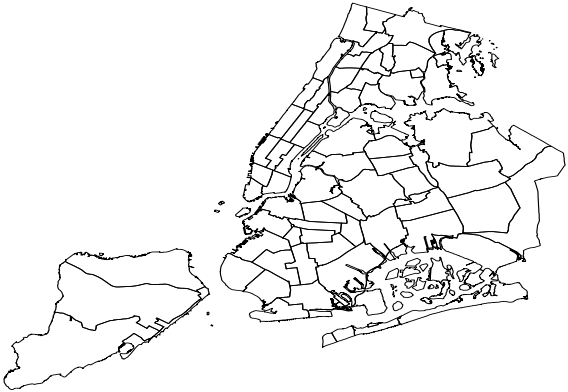}&\includegraphics[width=8cm,height=6cm]{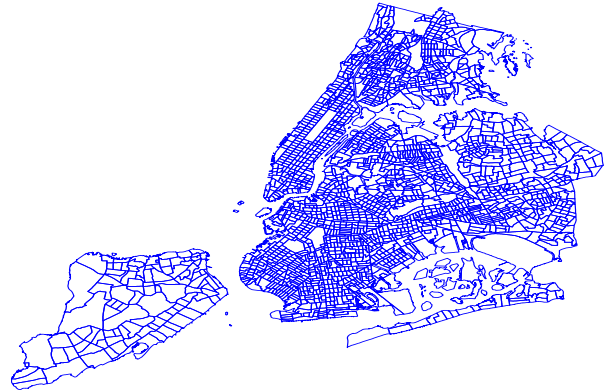}\\
    \includegraphics[width=8cm,height=6cm]{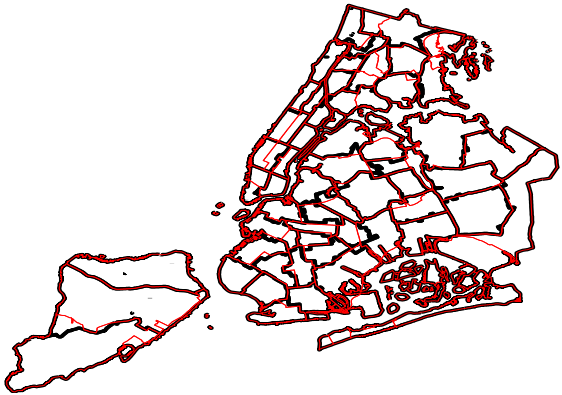}& \hfill
      \end{tabular}
      \end{center}
      \caption{In (a) we display the boundaries of the community districts, which is a target support for this example. In (b) we display the boundaries of the census tracts, which is the source support for this example. In (c), the overlap between PUMA (aggregate census tracts) and community districts are given. The black lines represent PUMA boundaries, and the red lines represent community district boundaries in NYC.}\label{fig:6}
      \end{figure}
  
  \newpage
    \begin{figure}[H]
    \begin{center}
    \begin{tabular}{c}
      \includegraphics[width=16cm,height=9cm]{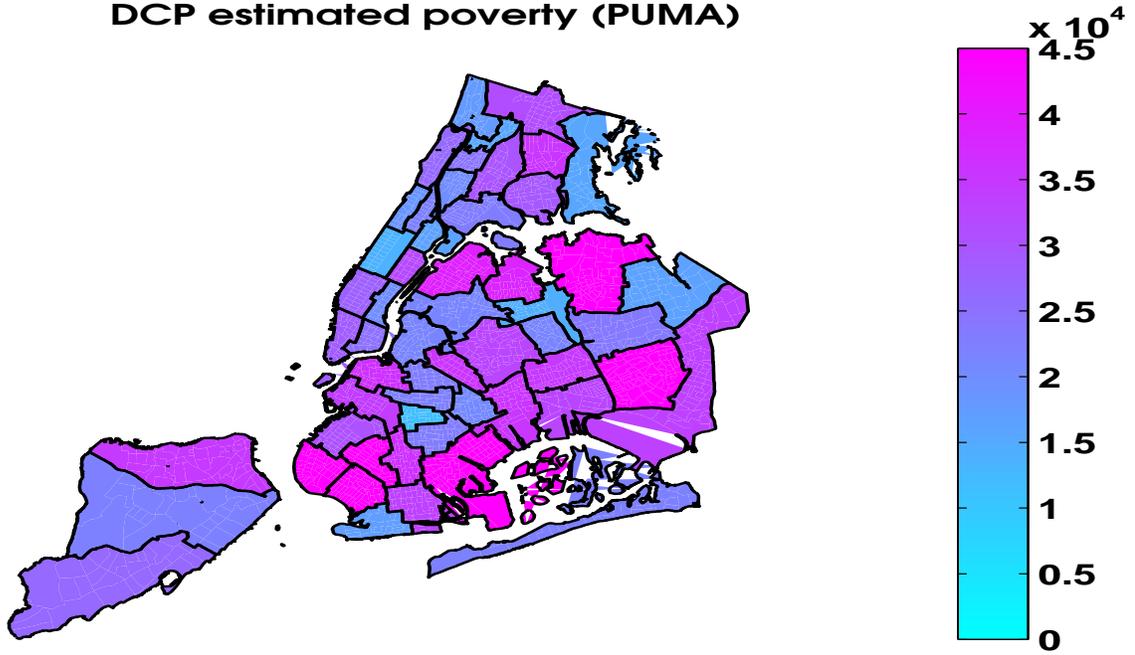}
    \end{tabular}
    \end{center}
    \caption{The New York City Department of City Planning approximations of poverty {(corresponding to the 2012 5-year period estimate)}. The values in this map are computed using simple areal interpolation. }\label{fig:6}
    \end{figure}
       
\newpage
  \begin{figure}[H]
  \begin{center}
  \begin{tabular}{cc}
  \includegraphics[width=6cm,height=6cm]{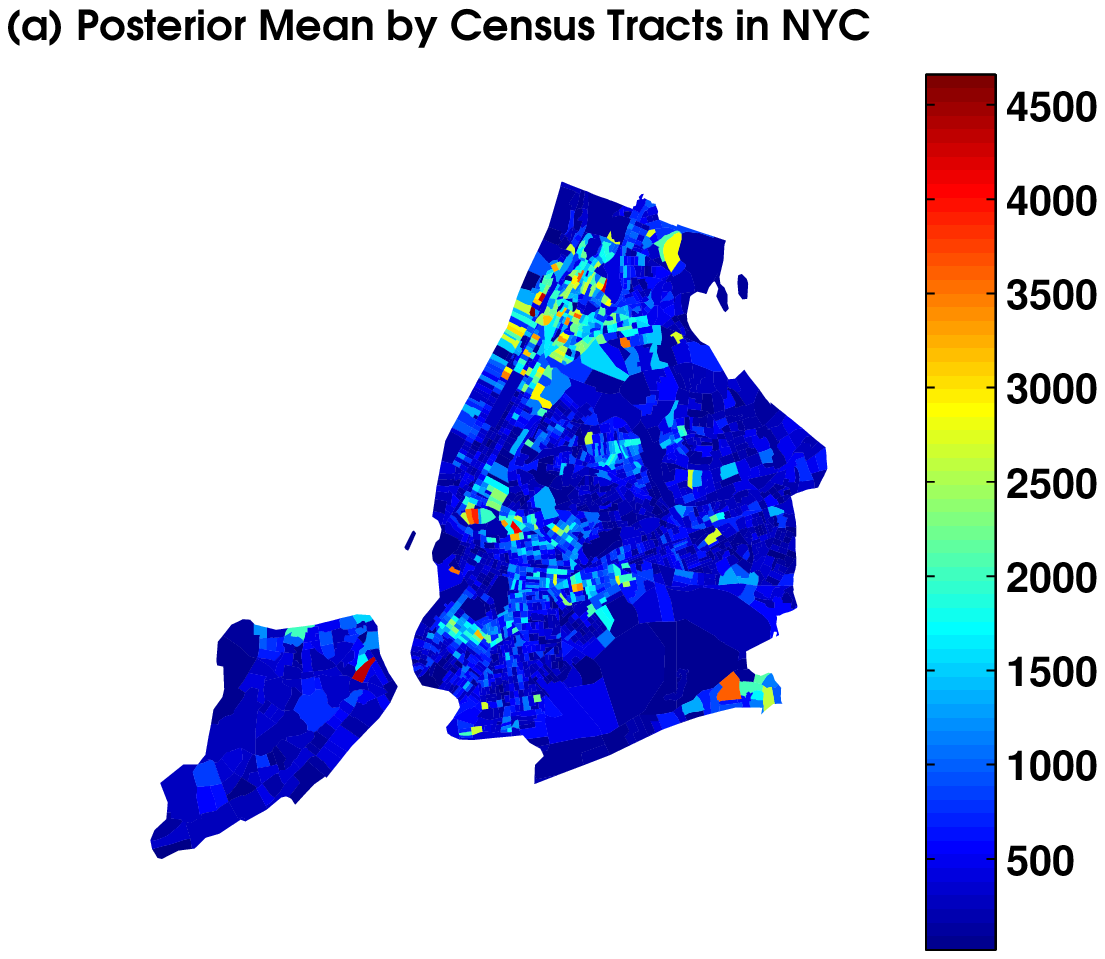}&\includegraphics[width=6cm,height=6cm]{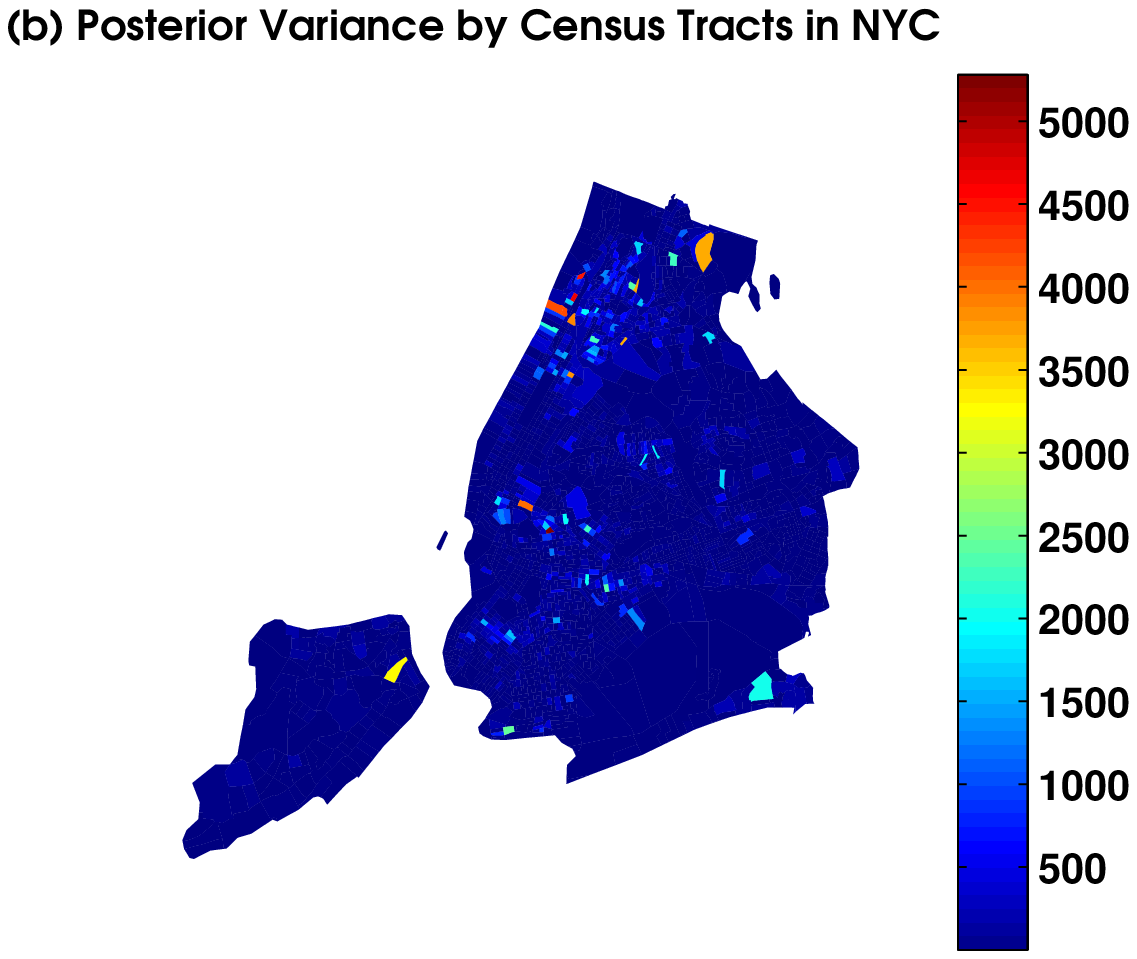}\\
  \includegraphics[width=6.5cm,height=6cm]{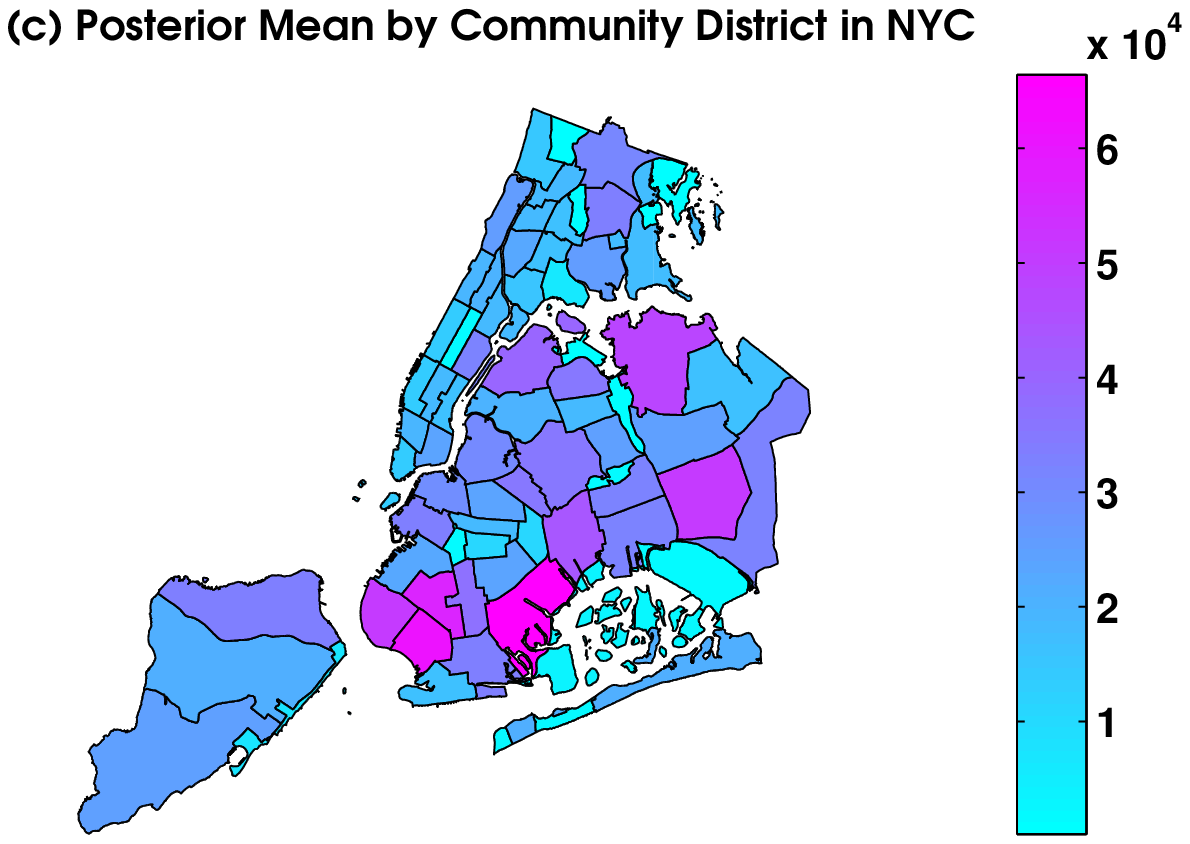}&\includegraphics[width=6.5cm,height=6cm]{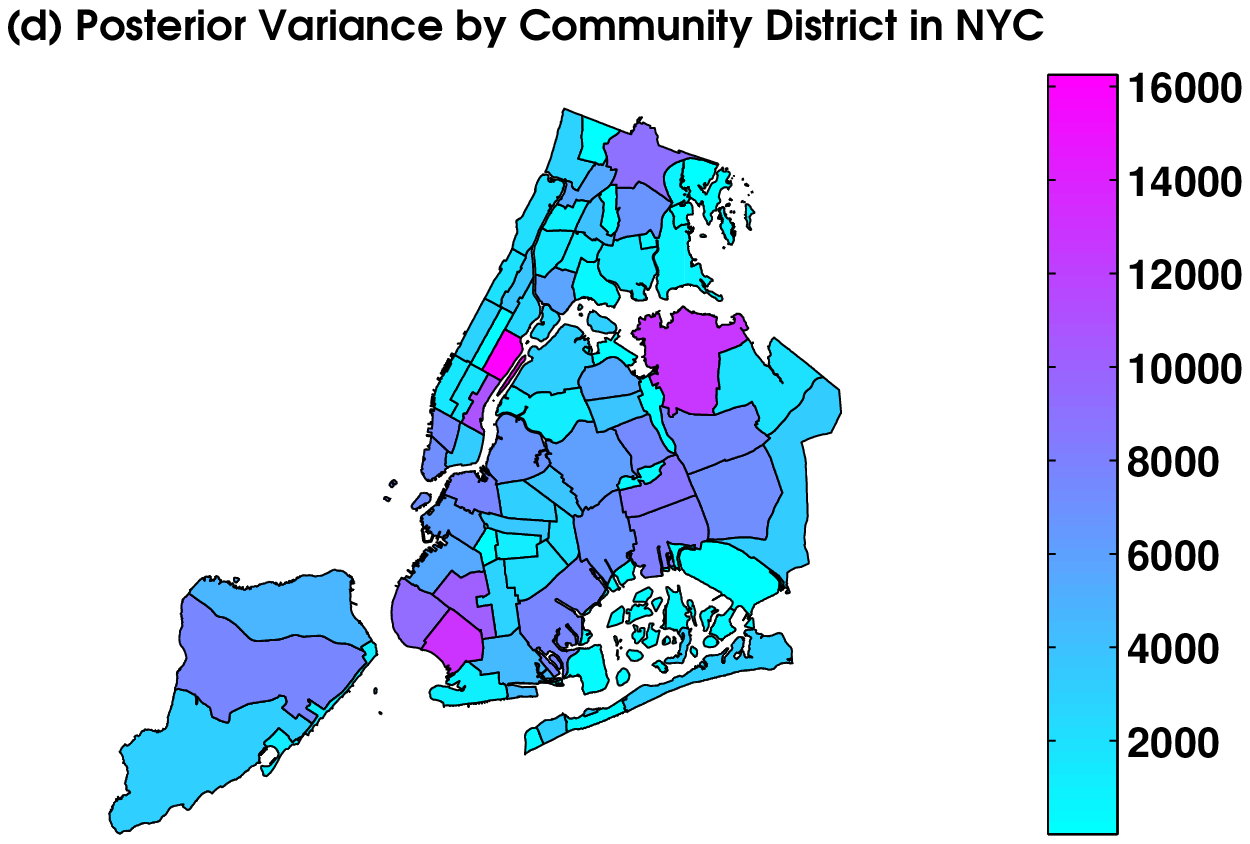}
  \end{tabular}
  \end{center}
  \caption{In (a) and (b) we present the posterior means of $\mu(\cdot)$ defined on census tracts (finest source support), and community districts (target support), respectively Each spatial support is indicated by the title headings. In (b) and (d) we present their respective posterior variances. Notice that the color-scales are different for each panel.}\label{fig:6}
  \end{figure}

\end{document}